\documentclass[a4paper, 11pt]{article}
\pdfoutput=1 % if your are submitting a pdflatex (i.e. if you have
\usepackage{jheppub} % for details on the use of the package, please
\usepackage[T1]{fontenc} % if needed
\usepackage{makecell}
\usepackage{graphicx}
\usepackage{subcaption}
\usepackage{tikz}
\usepackage{empheq}
\usepackage{subcaption}
\usepackage{hyperref}
\hypersetup{colorlinks=true
  ,urlcolor=blue
  ,anchorcolor=blue
  ,citecolor=blue
  ,filecolor=blue
  ,linkcolor=blue
  ,menucolor=blue
  ,pagecolor=blue
  ,linktocpage=true
  ,pdfproducer=medialab
  ,pdfa=true}

\newcommand*\widefbox[1]{\fbox{\hspace{1em}#1\hspace{1em}}}

\title{ \bf  Cardy Limits of   6d  Superconformal Theories }  % ( 6d Cardy fomulars for self-dual strings}

\author{Kimyeong Lee}
\author{and June Nahmgoong}

\affiliation{School of Physics, Korea Institute for Advanced Study, Seoul 02455, Korea}

\emailAdd{klee@kias.re.kr}
\emailAdd{junenahmgoong@gmail.com}

\abstract{ We explore the supersymmetric partition functions of 6d SCFTs on \(\mathbb{R}^4\times T^2\) with non-vanishing charges for compatible global symmetries.     We utilize the elliptic genera for self-dual strings   and compute the free energy of 6d SCFTs in the Cardy limit. For a 6d (2,0) theory on \(N\) M5-brane,  we obtain the free energy proportional to \(N^3\). We find that the origin of \(N^3\) comes from the condensation of the self-dual strings, whose total number is proportional to \({N^3-N \over 6}\). We further extend  our analysis to the general E-string theory and obtain its Cardy free energy. }

\preprint{KIAS-P20030}

\begin{document} 
\maketitle
\flushbottom

\section{Introduction and summary}
M5-branes have been the most mysterious object among the elements of M-theory \cite{Hull:1994ys, Witten:1995ex}. Compared to M2-branes \cite{Aharony:2008ug}, the physics of M5-branes is much less understood. The low energy dynamics of M5-branes are described by six-dimensional superconformal field theory \cite{Witten:1995zh, Seiberg:1996vs}. However,  this 6d theory of self-dual tensor fields lacks microscopic definition, and many of its detailed characteristics are mostly unknown.

\par

At the center of the puzzle, there is the \(N^3\) problem. For \(N\) M5-branes, it has been known that their entropy is proportional to \(N^3\) \cite{Klebanov:1996un} from the black \(p\)-brane solution. In the 6d SCFT side, the evidence of \(N^3\) scaling has been found in various observables, including the BPS string junction \cite{Bolognesi:2011rq}, the vacuum Casimir energy on \(S^5\) \cite{Kim:2012ava, Kallen:2012zn, Minahan:2013jwa, Kim:2013nva, Bobev:2015kza} and the 6d anomalies \cite{Freed:1998tg,Maxfield:2012aw,Ohmori:2014kda,Intriligator:2014eaa}. More direct evidence from the supersymmetric index itself was studied in \(\Sigma_{\mathfrak{g}_1}\times \Sigma_{\mathfrak{g}_2}\times T^2\) \cite{Hosseini:2018uzp}, \(\mathbb{R}^4\times T^2\) \cite{Kim:2011mv,Kim:2017zyo} and \(S^5\times S^1\) \cite{Nahmgoong:2019hko}. However, compared to the \(N^{3/2}\) scaling of M2-branes \cite{Drukker:2010nc, Herzog:2010hf}, the microscopic understanding of \(N^3\) degrees of freedom from 6d SCFTs needs further illustrations.

\par 

In this paper, we shall focus on the Cardy limit asymptotics of the 6d BPS index function on ${\mathbb R}^4\times T^2$ as a function of chemical potentials for compatible charges for the global symmetry.  This index function counts the Witten index of the massive BPS objects on the 6d theory on a circle. The Cardy limit refers to the large momenta limit, analogous to the original Cardy limit in 2d CFT \cite{Cardy:1986ie}. Therefore, the Cardy limit can be viewed as a high-energy limit in the BPS sector, where one can naturally expect a large number of microstate degeneracies. In the case of 6d SCFTs on $N$ M5 branes, the total number of the microscopic degrees is proportional to \(N^3\), and the resulting free energy should also be proportional to \(N^3\) in the large \(N\) limit. 

\par 

Until recently, it had been believed that the index does not capture such macroscopic free energy since the index counts BPS states with \((-1)^F\) \cite{Kinney:2005ej, DiPietro:2014bca}. However, it was pointed out that turning on the imaginary part of the chemical potentials can partially obstruct \((-1)^F\) the cancellation, and the macroscopic free energy can be obtained from the index \cite{Choi:2018vbz}. Furthermore, with the complexified chemical potentials, the Cardy limit of the superconformal indices account for the entropy of the dual \(AdS\) black holes in various dimensions \cite{Choi:2018hmj, Choi:2019miv, Kantor:2019lfo, Nahmgoong:2019hko, Choi:2019zpz, Nian:2019pxj}.

\par

Especially in six-dimensions, the Cardy formula in the complex chemical potential setting was studied in \cite{Nahmgoong:2019hko} with the background field method. Here, we shall take a different approach from \cite{Nahmgoong:2019hko} based on the localization formula of the elliptic genus. We will show that the final results are consistent with the background field method \cite{Nahmgoong:2019hko}, but our new approach reveals more microscopic details of the \(N^3\) free energy of 6d SCFTs.

\par

As a main observable, we consider the \(\mathbb{R}^4\times T^2\) index of 6d (2,0) \(A_{N-1}\) SCFT (with one additional Abelian tensor multiplet), which describes the low energy dynamics of \(N\) parallel M5-branes. On the field theory side, the 6d theory has a \(N-1\) dimensional moduli space called a tensor branch, which is parameterized by the VEVs of one of the five tensor multiplet scalars. There is a solitonic string called a self-dual string \cite{Howe:1997ue}. It is charged under the two-form field in the tensor multiplet, and its tension is proportional to the tensor VEVs. On the M-theory side, the distances between the M5-branes are proportional to the tensor VEVs. M2-branes can be suspended between the M5-branes, and the M2-brane ending on the M5-brane becomes the self-dual strings called M-strings \cite{Haghighat:2013gba}. 

\par 

As a constitutive element, the self-dual string is an essential object for studying 6d SCFTs. The \(\mathbb{R}^4\times T^2\) index of the 6d SCFT in the tensor branch counts the spectrum of the self-dual strings wrapping on \(T^2\). Recall that the self-dual strings have the tension, which is proportional to the VEVs of the scalars in the (2,0) tensor multiplet. For the index function, we turn on only one scalar field and keep the remaining four scalars to have zero expectation value. Therefore, the 6d index \(Z\) admits an expansion with respect to the string fugacity as follows,
\begin{align}
Z(\tau,z,{\bf v})= Z_0(\tau,z) \cdot \Bigr(1+ \sum_{\mathfrak{n}} Z_{\mathfrak{n}} (\tau,z) e^{-\mathfrak{n}\cdot {\bf v} }\Bigr)
\label{eq69_skim}
\end{align}
where \(\tau\) is the complex structure of \(T^2\), and \(z\) collectively denotes various chemical potentials. Here, \(\mathfrak{n}=(n_1,n_2,....,n_{N-1})\in \mathbb{Z}_{\geq 0}^{N-1}\) is the charge, or number of the self-dual strings and \({\bf v}=(v_1,v_2,...,v_{N-1})\in \mathbb{R}_{\geq 0}^{N-1}\) is the tensor VEV. For example, $v_i$  and $n_i$ denote  the relative distance and the number of M2 branes between $i$-th and $i+1$-th M5 branes along one transverse direction, say $x^6$ direction,  respectively. The overall factor \(Z_0(\tau,z)\) counts the pure momentum states from the \(N\) Abelian tensor multiplets. In the string fugacity expansion, the expansion coefficient \(Z_{\mathfrak{n}}\) is given by the elliptic genus of the self-dual string with charge \(\mathfrak{n}\). The elliptic genus of 6d self-dual string is a modular form of  weight zero and index \(\mathfrak{i}_{\bf n}(z)\). Under the \(SL(2,\mathbb{Z})\) transformation of \(T^2\), the elliptic genus transforms as follows,
\begin{align}
Z_{\mathfrak{n}}({a\tau+b\over c\tau+d},{z\over c\tau+d})=\varepsilon(a,b,c,d)\exp\Bigr[{-\pi i c \cdot \mathfrak{i}_{\mathfrak{n}}(z)\over c\tau+d} \Bigr]Z_{\mathfrak{n}}(\tau,z)
\label{eq72_sl2z}
\end{align}
where \(\begin{psmallmatrix} a & b \\ c & d\end{psmallmatrix}\in SL(2,\mathbb{Z})\), and \(\varepsilon\) is a \(z\)-independent phase. The modular index \(\mathfrak{i}_{\mathfrak{n}}(z)\) can be completely determined from the worldsheet chiral anomaly on the self-dual strings \cite{Benini:2013xpa, DelZotto:2017mee, Kim:2018gak}.

\par

In this paper, we compute the index of 6d SCFT in the Cardy limit where the spatial momenta become large. First, we take the two angular momenta $J_1, J_2$ on the spatial \(\mathbb{R}^4\) of M5 brane on the circle to be large.  Thus the corresponding chemical potentials, the Omega-deformation parameters \(\epsilon_{1,2}\), are taken to be small, which is the prepotential limit \cite{Nekrasov:2002qd}. Second, we take the KK momentum \(P\) on the spatial \(S^1\) in \(T^2\) to be large. The corresponding chemical potential limit is given by the limit where the \(T^2\) complex structure \(\tau\) is taken to be small. Therefore, in the Cardy limit, the modular property of the elliptic genus (\ref{eq72_sl2z}) becomes a useful tool to study the 6d index. Using the S-duality \(\tau\to -{1\over \tau}\), we obtain the asymptotic form of the self-dual string's elliptic genus in the Cardy limit. Then, we evaluate the elliptic genus summation (\ref{eq69_skim}) with the continuum approximation of the string number to obtain the Cardy free energy of the 6d SCFT.

\par

For 6d (2,0) SCFT on \(N\) M5-branes, we obtain the following free energy in the Cardy limit,
\begin{align}
\log Z=-{N^3\over 24}{m^2 (2\pi i -m)^2 +\mathcal{O}(\beta, \epsilon_{1,2})  \over \epsilon_1 \epsilon_2 \beta}
\label{eq84_cardyfree}
\end{align}
where \(\beta=-2\pi i \tau\). Here, \(m\) is the chemical potential for \(SU(2)_{L}\) which is the subset of \(SO(4)\) of  \(SO(5)\) R-symmetry of (2,0) theory and satisfies \({\rm Im}[m]  \in(0, 2\pi)\).  The scalar VEV with ${\bf v}$ breaks $SO(5)$ R-symmetry to $SO(4)=SU(2)_{R}\times SU(2)_{L}$. The remaining $SU(2)_{R}$ symmetry is locked to the spatial $SU(2)_{r}$, whose chemical potential is $\epsilon_+= (\epsilon_1+\epsilon_2)/2$. The \(N-1\) tensor VEVs are taken to be sufficiently small \(v_a\ll \mathcal{O}(\beta^{-1})\) for all $a$'s so that the 6d SCFT in the tensor branch acts like in the conformal phase. As we will see later, the $v_a$'s lose the role of chemical potential for the string numbers, which are determined by other chemical potentials. Then the free energy (\ref{eq84_cardyfree}) is explicitly proportional to \(N^3\) even at finite \(N\). Turning on the finite value of the imaginary part of \(m\) is crucial to obtain \(N^3\) free energy. The \((-1)^F\) cancellation of the index is maximally obstructed at \(m=\pi i\), and the index captures a macroscopic number of degrees of freedom.

\par 

If we move on the tensor branch moduli space by increasing the tensor VEV, we observe that the 6d SCFT undergoes phase transitions. When the tensor VEVs are larger than the critical value at the phase transition point, the free energy reduces to that of the \(N\) copies of the Abelian (2,0) tensor multiplets, and the 6d theory is in the confining phase.  However, when the tensor VEVs are smaller than the critical value, the non-Abelian contribution enhances the free energy to \(N^3\), and the 6d theory is in the deconfining phase. The free energy is maximized at the origin of the tensor branch, where the conformal symmetry is restored. 

\par

In the deconfining phase, we find that the number of the self-dual string has a non-zero expectation value. This condensation value of the self-dual string is proportional to a particular combinatoric factor, which can be interpreted with the bound state of W-boson and instantons in 5d SYM found in \cite{Kim:2011mv}. The total number of those degrees is given by \({N^3-N\over 6}\).

\par

 Our approach to the Cardy formula based on the elliptic genus sum can be applied to broader classes of 6d SCFTs with eight supercharges. Besides the (2,0) \(A\)-type theory, we extend our analysis to compute the Cardy free energy of the rank-\(N\) E-string theory, i.e., E-\(\text{M}^{N-1}\) string chain. It is a 6d (1,0) SCFT engineered from the worldvolume theory of \(N\) M5-branes probing a M9-brane \cite{Witten:1995gx, Ganor:1996mu,Seiberg:1996vs}. The elliptic genera of the E-strings were computed in \cite{Kim:2014dza, Cai:2014vka}. Using the modular property of the E-string elliptic genus, we compute the 6d free energy in the Cardy limit. 
 
\par

The rest of this paper is organized as follows. In section \ref{section2}, we study the 6d (2,0) SCFT on \(N\) M5-branes and compute its Cardy free energy on \(\mathbb{R}^4\times T^2\) with the continuum approximation of the elliptic genus summation. In section \ref{section_DSK}, we present the alternative approach to the Cardy formula based on the `S-duality kernel.' In section \ref{section3}, we extend our analysis to the 6d rank-\(N\) E-string theory. In section \ref{section4}, we re-derive the 6d free energies from the background field analysis using 6d 't Hooft anomalies. In section \ref{section5}, we conclude this paper with a few concluding remarks, including the possible implication to the gravity dual in \(AdS_7\).

\section{6d (2,0) \(A_{N-1}\) theory}
\label{section2}
In this section, we study the supersymmetric index of the 6d (2,0) \(A_{N-1}\)+1 free (2,0) tensor SCFT on \(\mathbb{R}^4\times T^2\). In subsection \ref{section21}, we briefly explain the 6d SCFT on the tensor branch and its index from the elliptic genera of the self-dual strings. In subsection \ref{section22}, we compute the free energy in the Cardy limit. The resulting Cardy free energy shows \(N^3\) growth at the conformal phase. In subsection \ref{section24}, we compute the asymptotic entropy of M5-branes from the Cardy free energy.

\subsection{\(\mathbb{R}^4\times T^2\) index}
\label{section21}

\begin{figure}[t!]
\centering
\begin{subfigure}{0.5\linewidth}
\centering
\includegraphics[width=\linewidth]{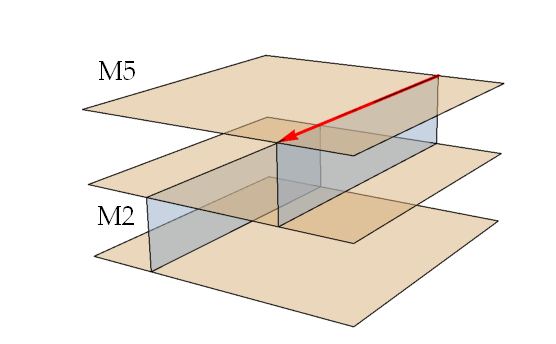}
\caption{M-theory picture of 6d (2,0) SCFT. The red arrow denotes the KK momentum along the compactified circle.}
\end{subfigure}\hfill
\begin{subfigure}{0.5\linewidth}
\centering
\includegraphics[width=\linewidth]{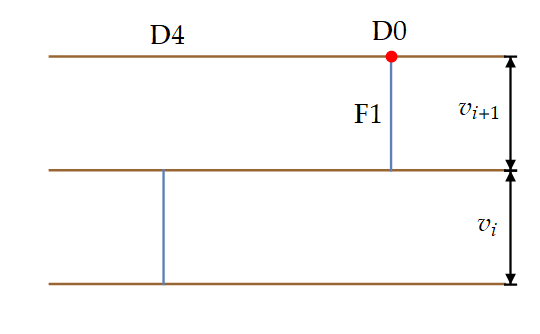}
\caption{IIA picture of 5d SYM.}
\end{subfigure}\hfill
\caption{Brane description of 6d SCFT/5d SYM on the tensor/Coulomb branch.}
\label{fig119}
\end{figure}

The 6d (2,0) \(A_{N-1}\) type SCFT is described by the non-Abelian (2,0) tensor multiplet which has the two-form tensor \(B_{\mu\nu}\) and five real scalars \(\Phi^I\) rotated by \(SO(5)\) R-symmetry. On the tensor branch, one of the tensor multiplet scalars has non-zero VEVs as \(\langle \Phi^5 \rangle =\sum_{a=1}^{N-1} V_a {\alpha}_a \). Here, \({\alpha}_a=\mathbf{e}_a-\mathbf{e}_{a+1}\) is the simple root of \(A_{N-1}\) and \(\mathbf{e}_a\) is the orthonormal basis of \(A_{N-1}\). The two form field \(B_{\mu\nu}\) has a field strength which is self-dual in 6d, and it is coupled to to the string solitons called the self-dual strings \cite{Howe:1997ue}. For (2,0) \(A\)-type theories, the self-dual string solitons are also called the M-strings \cite{Haghighat:2013gba}.

\par

In M-theory, the 6d (2,0) \(A_{N-1}\) SCFT can be constructed from the stack of \(N\) parallel M5-branes. Multiple M2-branes are suspended between the M5-branes, and the M2-brane ending on the M5-brane forms the M-string. The tensor VEV \(V_i\) parameterizes the distance between the \(i\)'th M5-brane and the \(i+1\)'th M5-branes, and the tension of the self-dual string suspended between the two M5-branes is proportional to \(V_i\). The center of mass degrees of freedom of \(N\) M5-branes are described by the free (2,0) Abelian tensor multiplet, and it is decoupled from the other non-Abelian degrees. In this paper, we shall consider (2,0) \(A_{N-1}\)+1 free tensor SCFT, which includes the center of mass degrees also.

\par 

Let us consider the 6d SCFT on \(\mathbb{R}^{1,4}\times S^1\) by compactifying a spatial direction on a circle with a radius \(R_1\). The dimensional reduction of the M5-M2 system yields a D4-F1-D0 system, and the number of D0-brane is proportional to the Kaluza-Klein momentum on \(S^1\). The worldvolume theory on D4-brane is the 5d \(U(N)\) maximal SYM with KK instantons \cite{Kim:2011mv}. The tensor VEVs become the Coulomb VEVs of the 5d vector multiplet, and they parameterize the 5d Coulomb branch moduli space. The brane descriptions of 6d SCFT and 5d SYM are shown in figure \ref{fig119}.

\par 

Now, let us define the BPS index of the circle compactified 6d SCFT. It has 16 supercharges, but we shall use (1,0) part of the supersymmetries to define the index. The eight (1,0) supercharges are given by \(Q^A_\alpha\) and \(Q^A_{\dot\alpha}\), where \(A=1,2\) is the doublet index for \(SU(2)_R\subset SO(5)\) R-symmetry, and \(\alpha,\dot{\alpha}=1,2\) are the doublet indicies for \(SU(2)_{l}\times SU(2)_{r}\subset SO(4)\) spatial rotation. The supersymmetry algebra contains the following anti-commutation relations \cite{Kim:2016usy},
\begin{align}
\{Q^A_\alpha,Q^B_\beta \}&=\epsilon^{AB} \epsilon_{\alpha\beta} \Bigr( E+{k\over R_1}+\sum_{a=1}^{N-1} V_a n_a \Bigr),\quad 
\{Q^A_{\dot \alpha},Q^B_{\dot \beta} \}=\epsilon^{AB} \epsilon_{\dot{\alpha}\dot{\beta} } \Bigr( E-{k\over R_1}-\sum_{a=1}^{N-1} V_a n_a \Bigr)
\nonumber \\
\{Q^A_\alpha, Q^B_{\dot\beta}\}&=\epsilon^{AB} (\sigma^\mu)_{\alpha\dot{\beta}}P_\mu.
\end{align}
Here, \(E\) is the energy, \(k\) is the KK-momentum, \(R_1\) is the radius of the compact spatial circle, \(P_\mu\) is the \(\mathbb{R}^4\) momentum, and \(n_a\)'s are the electric charges of \(U(1)^{N-1}\subset SU(N) \), which is the number of the self-dual strings. We shall study the index with a supercharge in \(Q^A_{\dot\alpha}\). Then, the BPS bound is given by \(E\geq {k\over R_1}+ \sum_a V_a n_a\), and the index preserves \({1\over 4}\) supersymmetries of the (2,0) theory. We put the 6d theory on the Euclidean temporal circle with radius \(R_0\) and define the index of the 6d (2,0) \(A_{N-1}\) SCFT on \(\mathbb{R}^4\times T^2\) as follows,
\begin{align}
Z=\text{Tr}\Bigr[ (-1)^F  e^{2\pi i \tau {R_1 E+k+R_1\sum V_a n_a\over 2} } 
e^{-2\pi i {\bar \tau} {R_1 E-k- R_1\sum V_a n_a\over 2} }
 e^{-\epsilon_1 J_1-\epsilon_2 J_2}e^{-(\epsilon_+ +m) Q_1-(\epsilon_+ -m) Q_2}\Bigr]
 \label{eq107_indexxx}
\end{align}
where \(\tau=iR_0/R_1\) is the complex structure of \(T^2\) and \(\epsilon_\pm ={\epsilon_1\pm \epsilon_2\over 2}\). Here, \(J_{1,2}\) are angular momenta of spatial \(SO(4)\), \(Q_{1,2}\) are charges of \(SO(5)\) R-symmetry and all of them are normalized to be \(\pm {1\over 2}\) for spinors. Since the BPS states satisfies \(E= {k\over R_1}+ \sum_a V_a n_a\), one can rewrite (\ref{eq107_indexxx}) as follows,
\begin{align}
Z=\text{Tr}\Bigr[ (-1)^F  e^{-\beta k} e^{-\mathfrak{n}\cdot \mathbf{v} }
 e^{-\epsilon_1 J_1-\epsilon_2 J_2}e^{-(\epsilon_+ +m)Q_1-(\epsilon_+ -m)Q_2}\Bigr]
\label{eq48_index}
\end{align}
where \(\mathfrak{n}=(n_1,...,n_{N-1})\) and \(\mathbf{v}=(v_1,...,v_{N-1})\). Also, we introduced a dimensionless tensor VEV \(v_a=2\pi R_0 V_a\) and a chemical potential \(\beta=-2\pi i \tau\). 
\par 
One way to study the \(\mathbb{R}^4\times T^2\) index is to compute the index of the 5d \(U(N)\) maximal SYM on \(\mathbb{R}^4\times S^1\) \cite{Kim:2011mv}. In the 5d viewpoint, the index receives contributions from the perturbative modes in SYM and non-perturbative instantons as follows,
\begin{align}
Z&=Z_\text{pert}(m,\epsilon_{1,2},\mathbf{v})\cdot \Bigr( 1+\sum_{k=1}^\infty q^k Z_k(m,\epsilon_{1,2},\mathbf{v})  \Bigr) 
\label{eq151_pertinst}
\end{align}
where \(q=e^{2\pi i \tau}\) is the instanton fugacity The perturbative partition function \(Z_\text{pert}\) is given as follows, \cite{Bullimore:2014awa, Bullimore:2014upa},
\begin{align}
Z_\text{pert}=\text{PE}\Bigr[ {\sinh{m\pm \epsilon_+\over 2 } \over \sinh{\epsilon_{1,2}\over 2}} \sum_{{\alpha}\in\Psi^+} e^{-{\alpha}\cdot \mathbf{v} } + {N\over 2}{\sinh{m\pm \epsilon_-\over 2 } \over \sinh{\epsilon_{1,2}\over 2}} \Bigr]
\label{eq170_pepert}
\end{align}
Here, we introduced \(\Psi^+=\{ \mathbf{e}_a-\mathbf{e}_b \ | \ 1\leq a<b\leq N \}\) as the set of the positive roots of \(SU(N)\) and \(\mathbf{v}=\sum_{a=1}^{N-1} v_a (\mathbf{e}_a-\mathbf{e}_{a+1})\) as a vector of tensor VEVs. Also we used a notation that \(f(x_{1,2})=f(x_1)f(x_2)\) and \(f(x\pm y)=f(x+y)f(x-y)\). The plethystic exponential is defined as \(\text{PE}[f(x)]=\exp[\sum_{n=1}^\infty {1\over n}f(nx)]\) where \(x\) is a chemical potential-like variable. The instanton partition function \(Z_k\) can be computed from the Witten index of \(k\) D0-branes on \(N\) D4-branes \cite{Kim:2011mv}, and it is the 5d uplift of the Nekrasov partition function \cite{Nekrasov:2002qd}. The explicit form of the \(U(N)\) instanton partition function is given as follows \cite{Kim:2011mv},
\begin{align}
Z_k=\sum_{Y_i;\sum_{a=1}^N |Y_a|=k} \prod_{a,b=1}^N \sum_{s\in Y_a} 
{\sinh{\hat{E}_{a,b}(s)+m-\epsilon_+\over 2 } \cdot \sinh{\hat{E}_{a,b}(s)-m-\epsilon_+\over 2 } 
\over 
\sinh{\hat{E}_{a,b}(s)\over 2} \cdot \sinh{\hat{E}_{a,b}(s)-2\epsilon_+\over 2}
}.
\end{align}
The summation is done over the `\(N\)-colored Young diagrams' \((Y_1,...,Y_{N})\) whose total size is given by the instanton number \(\sum_{a=1}^N |Y_a|=k\). Also, \(s=(p,q)\) denotes a single box at \((p,q)\) position in the Young diagram. The function \(\hat{E}_{a,b}(s)\) is defined as \(
\hat{E}_{a,b}(s)=v_a-v_b-\epsilon_1 H_a(s)+\epsilon_2(V_b(s)+1) \)  where \(H_a(s)\) is the distance from \(s\) to the right end of \(Y_a\), and \(V_b(s)\) is the distance from \(s\) to the bottom end of \(Y_b\). 
\par 
The other way to study the \(\mathbb{R}^4\times T^2\) index is to compute the elliptic genus of the M-strings \cite{Haghighat:2013gba}. In the elliptic genus method, the 6d index is expanded with respect to the string fugacity \(e^{-v_a}\) as follows,
\begin{align}
Z=[Z_{U(1)}]^{N}
\sum_{n_1=0}^\infty\sum_{n_2=0}^\infty...\sum_{n_{N-1}=0}^\infty Z_{\mathfrak{n}} e^{-\mathfrak{n}\cdot \mathbf{v} }
\label{eq65_index2}
\end{align}
Here, \(n_a\)'s are non-negative integers which are the charge, or the number of the self-dual strings. In the M-theory viewpoint, \(n_a\) is the number of the M2-branes suspended between the \(a\)'th and \(a+1\)'th M5-branes. The prefactor \(Z_{U(1)}\) is the Abelian partition function given as follows,
\begin{align}
Z_{U(1)}&=\text{PE}\Bigr[{\sinh{m\pm \epsilon_-\over 2}\over \sinh{\epsilon_{1,2}\over 2 }}\Bigr({1\over 2}+ {q\over 1-q} \Bigr)\Bigr].
\end{align}
The coefficient \(Z_{\mathfrak{n}} \) is the elliptic genus of the self-dual strings. It takes the following definition,
\begin{align}
Z_{\mathfrak{n}}=\text{Tr}\Bigr[ (-1)^F  e^{2\pi i \tau k} 
 e^{-\epsilon_1 J_1-\epsilon_2 J_2}e^{-(\epsilon_+ +m)Q_1-(\epsilon_+ -m)Q_2}\Bigr]
\end{align}
which is computed from the 2d quiver theory with \(U(n_1)\times U(n_2)\times...\times U(n_{N-1})\) gauge group \cite{Haghighat:2013tka}. The explicit form of \(Z_{\mathfrak{n}}\) is given as follows \cite{Haghighat:2013gba},
\begin{align}
Z_{\mathfrak{n}}=\sum_{|Y_a|=n_a} \prod_{a=1}^N \prod_{s\in Y_a} 
{
\theta_1(\tau|{m-\epsilon_- -E_{a,a+1}(s)\over 2\pi i}) \cdot 
\theta_1(\tau|{m+\epsilon_- + E_{a,a-1}(s)\over 2\pi i}) 
\over 
\theta_1({\tau|{\epsilon_1+E_{a,a}(s)\over 2\pi i})\cdot 
\theta_1(\tau|{\epsilon_2-E_{a,a}(s)\over 2\pi i}) } 
}
\label{eq78_ellipticgenus}
\end{align}
where we followed the notation in \cite{Kim:2015gha}. The summation is done over the `colored Young diagrams' \(\vec{Y}=(Y_1,...,Y_{N-1})\) whose size is given by \(|Y_a|=n_a\). By definition, \(Y_0\) and \(Y_N\) are empty Young diagrams and \(Y_{N+1}\equiv Y_1\). Also, \(s=(p,q)\) denotes a single box at \((p,q)\) position in the Young diagram. The function \(E_{a,b}(s)\) is defined as \(E_{a,b}(s)=(Y_{a,p}-q)\epsilon_1-(Y^T_{b,q}-p)\epsilon_2\) where \(Y_{a,p}\) is the length of the \(p\)'th row of \(Y_a\) and \(Y^T\) denotes the transposed Young diagram. We follow the notation of \cite{Kim:2015gha}. Lastly, our convention for the elliptic theta function is given as follows,
\begin{align}
\theta_1(\tau|z)&=-iq^{1\over 8} y^{1\over 2}\prod_{n=1}^\infty (1-q^n)(1-q^n y)(1-q^{n-1}y^{-1})
=-i\sum_{n\in\mathbb{Z} } (-1)^n y^{n+{1\over 2}} q^{{1\over 2}(n+{1\over 2})^2}
\label{eq83_theta}
\end{align}
where \(y=e^{2\pi i z}\). 
\par  
The elliptic genus is a weak Jacobi form of weight 0 and index \(\mathfrak{i}\), and the index \(\mathfrak{i}\) can be completely determined from the 2d chiral anomaly on the self-dual strings. In the case of the self-dual string in (2,0) \(A_{N-1}\) theory, one can explicitly check that the elliptic genus (\ref{eq78_ellipticgenus}) transforms as follows under the S-duality \cite{Kim:2017zyo},
\begin{align}
Z_{\mathfrak{n}}(\tau,m,\epsilon_{1,2})&=\exp\Bigr[ -{1\over 4\pi i \tau}\Bigr(\epsilon_1 \epsilon_2 \sum_{a,b=1}^{N-1} \Omega_{a,b} n_a n_b+2(m^2-\epsilon_+^2)\sum_{a=1}^{N-1} n_a \Bigr) \Bigr]
\nonumber \\
&\times Z_{\mathfrak{n}}\Bigr(-{1\over \tau},{m\over \tau},{\epsilon_{1,2}\over \tau}\Bigr)
\label{eq103_zn2}
\end{align}
where \(\Omega \) is the \(SU(N)\) Cartan matrix defined as \(\Omega_{a,b}=\{2\delta_{a,b}-\delta_{|a-b|,1} \ | \ 1\leq a,b\leq N-1 \}\).

\subsection{Free energy in the Cardy limit}
\label{section22}
In this subsection, we compute the free energy of the 6d (2,0) SCFT in the Cardy limit. On \(\mathbb{R}^4\times T^2\), the Cardy limit is defined as the large momenta limit where \(J_{1,2}\) and \(P\) are both large. In the canonical ensemble, we set the conjugate chemical potentials in (\ref{eq48_index}) to be small:
\begin{align}
\text{Cardy limit: }|\epsilon_{1,2}|\ll 1,\quad |\beta|\ll 1.
\label{eq230_highT}
\end{align}
Before computing the free energy, let us make a few comments on the leading behavior of the free energy in the Cardy limit.
\par 
First, \(\epsilon_{1,2}\) are the omega-deformation parameters of spatial \(\mathbb{R}^4\), and they regulate the effective volume of \(\mathbb{R}^4\) as \(\text{vol}(\mathbb{R}^4)\sim {1\over \epsilon_1 \epsilon_2}\). Hence small \(\epsilon_{1,2}\) limit corresponds to the thermodynamic limit where the free energy is proportional to the volume, i.e. \(\lim_{\epsilon_{1,2}\to 0}\log Z =\mathcal{O}({1\over \epsilon_1 \epsilon_2})\). In this limit, the leading behavior can be studied from the Seiberg-Witten prepotential of the Nekrasov partition function \cite{Nekrasov:2002qd}. 
\par 
Second, \(\beta={2\pi R_0\over R_1}\) is related to the complex structure of \(T^2\) made of the spatial circle with radius \(R_1\) and the temporal circle with radius \(R_0\). Since the volume of the spatial circle is inversely proportional to \(\beta\), we expect that \(\lim_{\beta\to 0}\log Z =\mathcal{O}({1\over \beta})\). However, it is much more challenging to obtain the free energy in \(\beta\to 0\) limit than \(\epsilon_{1,2}\to 0\) limit. The main reason is that, in small \(\beta\) limit, the KK instantons become massless, and the contribution from infinitely many instantons has to be resumed in the full partition function (\ref{eq151_pertinst}). In this paper, we will circumvent this problem by using the \(SL(2,\mathbb{Z})\) property of the elliptic genus. As a result, we will see that the leading free energy in the Cardy limit is of order \(\mathcal{O}({1\over \epsilon_1 \epsilon_2 \beta})\). We shall focus on the leading order, and the relative scaling between \(\epsilon_{1,2}\) and \(\beta\) is not important. 
\par 
The chemical potentials \(\epsilon_{1,2}\), \(\beta\), and \(m\) can have generic complex values except that \(\beta\) is constrained by \(\text{Re}[\beta]>0\). In the rest of this paper, for the simplicity, we shall set small parameters \(\epsilon_{1,2}\) and \(\beta\) to be purely real as follows,
\begin{align}
\epsilon_1>0,\quad \epsilon_2<0,\quad \beta>0.
\label{eq278_chemset}
\end{align}
Note that the signs of \(\epsilon_{1,2}\) are opposite. In thi setting, we choose \(e^{-\epsilon_1}\), \(e^{+\epsilon_2}\), and \(e^{-\beta}\) to be the expansion parameters of the index. This choice of the fugacities is also analogous to the topologial string theory set-up where \(e^{-\epsilon_1}\) and \(e^{+\epsilon_2}\) become the equivariant parameters of the refined topological partition function \cite{Iqbal:2007ii}. 
\par 
The only chemical potential that can have \(\mathcal{O}(1)\) value in the Cardy limit is \(m\). It is the R-charge chemical potential conjugate to \(Q_1-Q_2\), and we take \(m\) to be purely imaginary,
\begin{align}
\text{Re}[m]=0.
\label{eq238_rem}
\end{align}
Lastly, on the tensor branch, we are interested in the free energy in the conformal phase where \(v\to 0\). However, one may not immediately insert \(v_a=0\) in the \(\mathbb{R}^4\times T^2\) index since there is a divergence of the instanton partition function \cite{Kim:2011mv} at \(v_a=0\). Although this divergence is not captured from the elliptic genus expansion, we should be more careful about going to the conformal phase. Therefore, we shall keep \(v_a\) to be a real number of order \(\mathcal{O}(\beta^{-1})\). In the end, we will approach to the conformal phase by setting \(\beta v_a\to 0\).
\par
Now, let us compute the free energy of 6d theory on \(\mathbb{R}^4\times T^2\) from the elliptic genus of the self-dual strings. Using the S-duality property of the elliptic genus (\ref{eq103_zn2}), we can rewrite the \(\mathbb{R}^4\times T^2\) index (\ref{eq65_index2}) with the S-dualized elliptic genus as follows, 
\begin{align}
Z
&=[Z_{U(1)} ]^N   \sum_{\mathfrak{n}} Z_{\mathfrak{n}} e^{-\mathfrak{n}\cdot \mathbf{v} }
\nonumber \\
&= [Z_{U(1)}]^N    \sum_{\mathfrak{n} }  \exp\Bigr[ {\epsilon_1 \epsilon_2\over 2\beta }\sum_{a,b} \Omega_{a,b} n_a n_b
+\sum_a ( {m^2-\epsilon_+^2\over \beta }-v_a) n_a  \Bigr]
 Z_{\mathfrak{n}}^D
\label{eq143_fullindex}
\end{align}
where \(Z_{\mathfrak{n}}^D=Z_{\mathfrak{n}}(-{1\over \tau},{m\over \tau},{\epsilon_{1,2}\over \tau})\) is the dual elliptic genus. We set \(m\) to be a purely imaginary number, and it is periodic under \(m\sim m+2\pi i\). Here, we shall perform the computation in the following range of \(m\) which we call a `canonical chamber,'
\begin{align}
0<\text{Im}[m  ]<2\pi.
\label{eq153_CANONICAL}
\end{align}
The above chamber is chosen for the convenience of the computation, and we will discuss the free energy outside the canonical chamber later.
\par 
In the canonical chamber, the dual elliptic genus \(Z_{\mathfrak{n}}^D\) has the simple asymptotic form in the Cardy limit. Recall that the elliptic genus (\ref{eq78_ellipticgenus}) can be written in terms of the theta functions follows,
\begin{align}
Z_{\mathfrak{n}}^D=\sum_{|Y_a|=n_a} \prod_{a=1}^N \prod_{s\in Y_a} 
{
\theta_1(-{1\over \tau} |{m-\epsilon_- - E_{a,a+1}(s)\over 2\pi i \tau}) \cdot 
\theta_1(-{1\over \tau}|{m+\epsilon_- + E_{a,a-1}(s)\over 2\pi i \tau}) 
\over 
\theta_1({-{1\over \tau}|{\epsilon_1+E_{a,a}(s)\over 2\pi i \tau})\cdot 
\theta_1(-{1\over \tau}|{\epsilon_2-E_{a,a}(s)\over 2\pi i \tau}) } 
}.
\label{eq322_sde}
\end{align}
In the canonical chamber, all theta functions \(\theta_1(-{1\over \tau}|z)\) in (\ref{eq322_sde}) has argument such that \(1\leq |y|<(q^D)^{-1}\) where \(y=e^{2\pi iz}\) and \(q^D=e^{-2\pi i /\tau}\). Therefore, from the definition of the theta function (\ref{eq83_theta}), all of them can be approximated as trigonometric functions as follows,
\begin{align}
Z_{\mathfrak{n}}^D
&=\sum_{|Y_a|=n_a} \prod_{a=1}^N \prod_{s\in Y_a} 
{
\sinh({m-\epsilon_- -E_{a,a+1}(s)\over  2\tau}) \cdot 
\sinh({m+\epsilon_- +E_{a,a-1}(s)\over  2\tau}) 
\over 
\sinh({ \epsilon_1+ E_{a,a}(s)\over 2\tau})\cdot 
\sinh({\epsilon_2 -E_{a,a}(s)\over  2\tau}) 
}\Bigr(1+\mathcal{O}(q^D) \Bigr).
\label{eq300_dualellip}
\end{align}
Since \({\epsilon_{1,2}\over \tau}\) is a pure imaginary number and \({m\over \tau}\) is a pure real number, each \(\sinh\) in the numerator of (\ref{eq300_dualellip}) has an exponentially large factor \(e^{m\over \tau}\). As a result, the dual elliptic genus has the following asymptotic form in the Cardy limit,
\begin{align}
Z_{\mathfrak{n}}^D
=\exp\Bigr[ {1\over \beta}\Bigr(-2\pi i m +\mathcal{O}(\epsilon_{1,2}) \Bigr)\sum_{a=1}^{N-1} n_a+\mathcal{O}(\beta^0) \Bigr]. 
\label{eq294_dualellip}
\end{align}
Inserting the asymptotic form (\ref{eq294_dualellip}) to (\ref{eq143_fullindex}), the elliptic genus summation of the tensor branch index can be written as follows,
\begin{align}
&Z= [Z_{U(1)}]^N    
\sum_{\mathfrak{n} }  \exp\Bigr[ {\epsilon_1 \epsilon_2\over 2\beta }\sum_{a,b} \Omega_{a,b} n_a n_b
+\sum_a \Bigr( {m(m-2\pi i) -\beta v_a+\mathcal{O}(\epsilon_{1,2})\over \beta } \Bigr) n_a+\mathcal{O}(\beta^0)   \Bigr] . 
\label{eq278_tbi}
\end{align}
One may worry that \(\mathcal{O}(\beta^0) \) term can spoil the above summation structure if the subleading correction \(\mathcal{O}(\beta^0) \) grows too rapidly when \(\mathfrak{n}\) grows. However, in \(\mathcal{O}(\beta^0)\), the maximally growing factor in the large \(\mathfrak{n}\) is the degeneracy of the colored Young diagrams in (\ref{eq300_dualellip}), which gives \(\mathcal{O}(n^{1\over 2})\). It is much more subleading than \(\mathcal{O}(n^2)\) and \(\mathcal{O}(n^1)\) terms in the exponent. Hence, we can safely ignore the subleading terms in the Cardy limit also when \(n_a\)'s are large.
\par
\begin{figure}
\centering
\includegraphics[width=0.6\columnwidth]{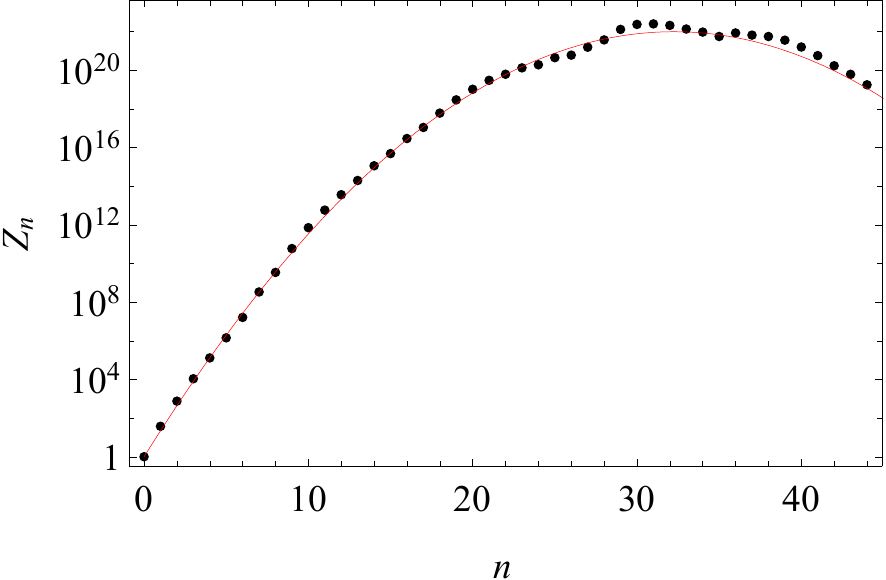}
\caption{Elliptic genera of the self-dual strings in (2,0) \(A_1\) SCFT computed up to \(n=45\) charges. The chemical potentials are \(\epsilon_1=0.433\), \(\epsilon_2=-0.354\), \(\beta=\pi\), and \(m=\pi i\). Red line is the Gaussian approximation given by \( Z_n=\exp\Bigr[{\epsilon_1 \epsilon_2\over\beta}n^2+{m(m-2\pi i)\over \beta}n\Bigr]\). Even though \(\epsilon_{1,2}\) and \(\beta\) are finite, the asymptotic form in the Cardy limit well approximates the elliptic genus.}
\label{fig320}
\end{figure}

 
Inside the summation (\ref{eq278_tbi}), the summand takes the form of the Gaussian on the \(N-1\) dimensional space of the string number \(n_a\)'s. The eigenvalues of the matrix \({\epsilon_1 \epsilon_2\over 2\beta}\Omega\)  are all negative, and therefore, the summation is convergent. In figure \ref{fig320}, we present a numerical computation of the elliptic genus of (2,0) \(A_1\) theory, and it shows that the elliptic genus can be well-approximated with (\ref{eq278_tbi}). 
\par 
Now, we evaluate the summation in (\ref{eq278_tbi}) using the continuum approximation  of the string charge \(n_a \in \mathbb{Z}_{\geq 0}\). Let us define a variables \(x_a\) as follows,
\begin{align}
x_a\equiv -\epsilon_1 \epsilon_2 n_a.
\end{align}
Since \(0<-\epsilon_1 \epsilon_2\ll 1\), we can regard \(x_a\)'s as a continuous variables in \((0,\infty)\). By changing variables from \(n_a\)'s to \(x_a\)'s, one can see that the leading free energy is the order of \(\mathcal{O}({1\over \epsilon_1 \epsilon_2\beta})\). At the leading order, the index (\ref{eq278_tbi}) is simplified to the following Gaussian integral,
\begin{align}
&Z=[Z_{U(1)}]^N
\nonumber \\
&\times \int_0^\infty [\prod_a {dx_a\over |\epsilon_1 \epsilon_2|} ]
\exp\Bigr[{1\over \epsilon_1 \epsilon_2\beta}\Bigr( {1\over 2}\sum_{a,b}\Omega_{a,b}x_a x_b
+\sum_a [m(2\pi i-m)+\beta v_a ]x_a +\mathcal{O}(\epsilon_{1,2}) \Bigr)+\mathcal{O}(\beta^0)  \Bigr].
\label{eq374_contiapprox}
\end{align}
Here, note that (\ref{eq374_contiapprox}) cannot be used to accurately approximate the index itself due to the subleading correction in the exponent. However, since we are focusing the free energy \(\log Z\), (\ref{eq374_contiapprox}) can be used to obtain leading free energy in the Cardy limit.
\par 
 We can compute the leading free energy in the Cardy limit from the saddle point approximation of the integral (\ref{eq374_contiapprox}). In this case, the integrand takes the form of the Gaussian. The saddle point becomes the peak where the integrand is maximized on \(\mathbb{R}^{N-1}\). However, note that the integral is performed over the positive real space \(\mathbb{R}_{\geq 0}^{N-1}\). Therefore, one can approximate the integral to the value at the saddle point only when the saddle point is located in \(\mathbb{R}^{N-1}_{\geq 0}\). Otherwise, one should maximize the integrand along the boundary of \(\mathbb{R}^{N-1}_{\geq 0}\).
 \par 
For simplicity, let us consider the case when all tensor VEVs are equal, i.e., \(v_a\equiv v\). Then, one can easily find that the integrand is maximized on \(\mathbb{R}^{N-1}_{\geq 0}\) at the following point,
 \begin{align}
 \hat{x}_a&={a(N-a)\over 2}[m(m-2\pi i)-\beta v]+\mathcal{O}(\epsilon_{1,2}),& &0<v<{m(m-2\pi i)\over \beta}
 \nonumber \\
 &=\mathcal{O}(\epsilon_{1,2}),& &{m(m-2\pi i)\over \beta}<v.
 \label{eq327_saddel}
 \end{align}
Here, we used a relation \(\sum_b \Omega^{-1}_{a,b}={a(N-a)\over 2}\) for \(A_{N-1}\) Cartan matrix. Note that the first line of (\ref{eq327_saddel}) is the saddle point of the integral which is valid only when \(v\) is smaller than \({m(m-2\pi i )\over \beta}\). Otherwise, the saddle point lies outside of the integral range, and the integrand is maximized at the boundary point \(x_a=0\). Now, it is straightforward to obtain the Cardy free energy from (\ref{eq327_saddel}), and the results are given as follows,
\begin{align}
\log Z&=-{N\over 24}{m^2(2\pi i-m)^2\over \epsilon_1 \epsilon_2 \beta}-{N^3-N\over 24}{[m(m-2\pi i)-\beta v]^2\over \epsilon_1 \epsilon_2 \beta}+{\mathcal{O}(\epsilon_{1,2})\over \epsilon_1 \epsilon_2 \beta},& &0<v<{m(m-2\pi i)\over \beta}
\nonumber \\
&=-{N\over 24}{m^2(2\pi i-m)^2\over \epsilon_1 \epsilon_2 \beta}+{\mathcal{O}(\epsilon_{1,2})\over \epsilon_1 \epsilon_2 \beta},& &{m(m-2\pi i)\over \beta}<v.
\label{eq332_deconf_free}
\end{align}
Here, we used the Abelian Cardy formula \(\log Z_{U(1)}\simeq -{1\over 24}{m^2(2\pi i -m)^2 \over \epsilon_1 \epsilon_2 \beta}\) obtained in \cite{Kim:2017zyo}. Also, we used a group theory identity \(\sum_{a,b}\Omega^{-1}_{a,b}={N^3-N\over 12}\) for \(A_{N-1}\) Cartan matrix. As we can see, the non-Abelian part of the free energy contains \(N^3\) factor as expected for 6d SCFTs. Taking the conformal phase limit \(v\to 0 \) is smooth, and we finally obtain the following Cardy formula for 6d (2,0) \(A_{N-1}\) theory,
\begin{align}
\boxed{\log Z=-{N^3\over 24}{m^2(2\pi i -m)^2+\mathcal{O}(\beta, \epsilon_{1,2})\over \epsilon_1 \epsilon_2 \beta} .}
\label{eq343_cardy}
\end{align}
The above \(\mathcal{O}(N^3)\) free energy of the 6d SCFTs on \(N\) M5-branes (\ref{eq343_cardy}), as far as we know, is the first microscopic computation from the supersymmetric index without the Casimir factor. Note that the non-Abelian contribution proportional to \(N^3-N\) and the Abelian contribution proportional to \(N\) combine to \(N^3\) even at the finite \(N\). Turning on the imaginary part of \(m\) is crucial to obtain the above macroscopic free energy, since the phase factor \(e^{-m(Q_1-Q_2)}\) in the index (\ref{eq48_index}) can obstruct \((-1)^F\) cancellation, especially at \(m=\pi i\). 
\par
\begin{figure}[t!]
\centering
\includegraphics[width=0.7\columnwidth]{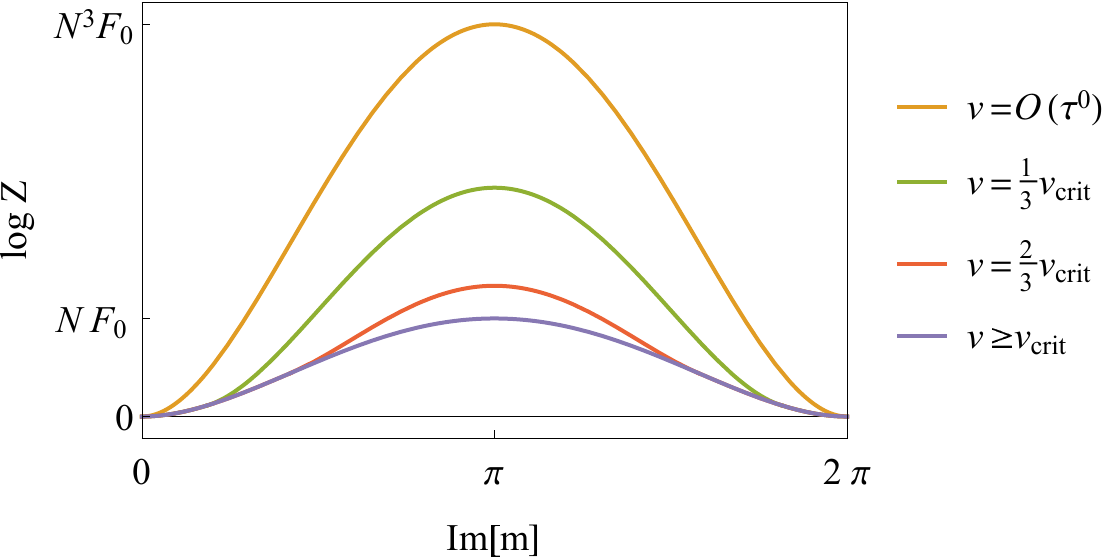}
\caption{The free energy normalized with \(F_0=-{\pi^4\over 24\epsilon_1 \epsilon_2\beta }\). As we decrease \(v\), the free energy enhances from \(N^1\) to \(N^3\). \(v_\text{crit}={\pi^2\over \beta}\) is the critical value where the phase transition can happen.}
\label{fig331}
\end{figure}

 
One interesting point of the Cardy free energy is that there are two phases on the tensor branch. The first one is the `deconfining' phase \(0<v<{m(m-2\pi i)\over \beta}\) where the free energy (\ref{eq332_deconf_free}) sees the non-Abelian \(N^3\) enhancement. The second one is the `confining' phase \({m(m-2\pi i )\over \beta}<v\) where the free energy remains Abelian. Also, \(v_\text{crit}={\pi^2\over \beta}\) becomes the critical value for the phase transition, and the free energy cannot deconfine if \(v>v_\text{crit}\). The phases of free energy are illustrated in figure \ref{fig331}. On the tensor branch, the non-Abelian tensor in 6d spontaneously breaks down into the \(N-1\) Abelian tensors due to non-zero tensor VEVs. Therefore, it is natural to expect that the non-Abelian structure arises when the tensor VEVs are small, as we observed.
\par 
Besides the free energy, our analysis also sheds light on the \(N^3\) degrees of freedom in 6d. Recall that we computed the 6d free energy by summing over the elliptic genera of the self-dual strings. Then, one can define the `expectation value' of the self-dual string's number as follows,
\begin{align}
\langle n_a \rangle={\sum_{\mathfrak{n}} n_a Z_{\mathfrak{n}}e^{-\mathfrak{n}\cdot v}\over \sum_{\mathfrak{n}}Z_{\mathfrak{n}}e^{-\mathfrak{n}\cdot v}}=-{\partial\over \partial v_a}\log Z.
\end{align}
Also, it can be computed from the saddle point approximation of the free energy by generalizing (\ref{eq332_deconf_free}) to the non-equal tensor VEV setting. If we set \(v_a\)'s to be unequal but satisfy \(0<v_a<{m(m-2\pi i)\over \beta}\), the free energy (\ref{eq332_deconf_free}) is generalized as follows,
\begin{align}
\log Z=-{N\over 24}{m^2(2\pi i-m)^2\over \epsilon_1 \epsilon_2 \beta}
-\sum_{a,b}^{N-1} \Omega_{a,b}^{-1}{ [m(m-2\pi i)-\beta v_a][m(m-2\pi i)-\beta v_b]\over 2\epsilon_1 \epsilon_2 \beta}
\label{eq367_freeenergy}
\end{align}
where we ignored subleading corrections which are unimportant. As a result, at the conformal phase \( v_a\to 0\), one obtains the following expectation value,
\begin{align}
\langle n_a \rangle = a(N-a){m(2\pi i-m)\over 2\epsilon_1 \epsilon_2}.
\label{eq364_stringcondense}
\end{align}
We can see that the self-dual strings condensate to a large number of order \(\mathcal{O}({1\over \epsilon_1 \epsilon_2})\) in the Cardy limit. Here, the combinatoric factor \(n_a\propto a(N-a)\) in (\ref{eq364_stringcondense}) is identical to the Weyl vector of \(A_{N-1}\) Lie algebra. More precisely, by expanding the Weyl vector in terms of the simple roots, its coefficient is given by \(a(N-a)\) as follows,
\begin{align}
\sum_{\alpha\in\Psi^+}\alpha =\sum_{a<b} (\mathbf{e}_a-\mathbf{e}_b) =\sum_{a=1}^{N-1} a(N-a)\cdot (\mathbf{e}_a-\mathbf{e}_{a+1} ).
\end{align}
\par
\begin{figure}[t!]
\centering
\includegraphics[width=0.6\columnwidth]{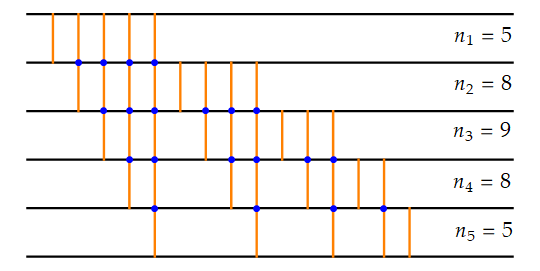}
\caption{6d (2,0) theory on \(N=6\) M5-branes. Orange lines denote the W-bosons, and the blue dots denote the KK instantons. Considering their bound states, the number of the independent degrees between \(a\)'th and \(a+1\)'th is proportional to \(a(N-a)\). }
\label{fig372}
\end{figure}

 
Physically, the factor \(a(N-a)\) can also be interpreted as the bound state of the self-dual strings and KK momentum \cite{Kim:2011mv}. In \cite{Kim:2011mv}, the authors studied 6d \(\mathbb{R}^4\times T^2\) index from the 5d instanton partition function on \(\mathbb{R}^4\times S^1\). By investigating the charged instanton sector, they observed that the single-particle index acquires a universal factor when the W-boson crosses a D4-brane. This is strong evidence that W-bosons and instantons form nontrivial bound states. The independent degrees of such bound states is \({}_N C_3={N(N-1)(N-2)\over 6}\), and together with \({}_N C_2={N(N-1)\over 2}\) number of W-bosons, the total number of the non-Abelian degrees becomes \({N^3-N\over 6}\). Now, let us return to our result (\ref{eq364_stringcondense}) and the factor \(a(N-a)\). See figure \ref{fig372} for the illustration. In 6d, the bound states of W-bosons and instantons are uplifted to the bound states of the self-dual strings and the KK momentum. Since they form bound states, the number of the independent degrees of freedom between \(a\)'th and \(a+1\)'th D4-branes is \(a(N-a)\) which exactly matches with the expectation value of the self-dual string condensation in (\ref{eq364_stringcondense}). Moreover, summing over \(\langle n_a \rangle\) yields
\begin{align}
\sum_{a=1}^{N-1} a(N-a)={N^3-N\over 6}
\end{align}
which is the same with the total number of non-Abelian degrees found in \cite{Kim:2011mv}.
\par
\begin{figure}
\centering
\includegraphics[width=0.5\columnwidth]{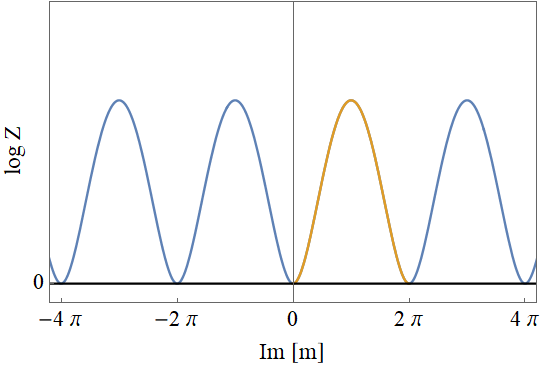}
\caption{Cardy free energy of the 6d (2,0) theory at the conformal phase. The yellow line is the free energy inside the canonical chamber where the expression (\ref{eq343_cardy}) is valid. }
\label{eq417}
\end{figure}

 
Up to now, the previous analysis was performed in the canonical chamber (\ref{eq153_CANONICAL}). Now, let us discuss the free energy outside of the canonical chamber. Here, we will set \(m\) to be a purely imaginary number in the following range,
\begin{align}
2\pi p< \text{Im}[m]<2\pi(p+1),\quad p\in\mathbb{Z}.
\label{eq423_outside}
\end{align}
We should re-compute the Cardy limit asymptotics of the dual elliptic genus on the above chamber.  Recall the theta function expression of the elliptic genus,
\begin{align}
Z_{\mathfrak{n}}^D
=\sum_{|Y_a|=n_a} \prod_{a=1}^N \prod_{s\in Y_a} 
{
\theta_1(-{1\over \tau} |{m-\epsilon_--E_{a,a+1}(s)\over 2\pi i \tau}) \cdot 
\theta_1(-{1\over \tau}|{m+\epsilon_-+E_{a,a-1}(s)\over 2\pi i \tau}) 
\over 
\theta_1({-{1\over \tau}|{\epsilon_1+E_{a,a}(s)\over 2\pi i \tau})\cdot 
\theta_1(-{1\over \tau}|{\epsilon_2-E_{a,a}(s)\over 2\pi i \tau}) } }.
\label{eq389_outsi}
\end{align}
The theta function \(\theta_1(-{1\over \tau}|z)\) in the numerator satisfies that \((q^D)^{-p}<|y|<(q^D)^{-p-1}\) where \(y=e^{2\pi i z}\) and \(q^D=e^{-2\pi i/\tau}\). Then, those theta functions can be approximated as follows,
\begin{align}
\theta_1(-{1\over \tau}|z)&=-i\sum_{l=-\infty}^\infty (-1)^l y^{l+{1\over 2}} (q^D)^{{1\over 2}(l+{1\over 2})^2} 
\simeq -i (-1)^p y^{p+{1\over 2}} (q^D)^{{1\over 2}(p+{1\over 2})^2}
\label{eq395_dfkbdm}
\end{align}
On the other hands, the theta functions \(\theta_1(-{1\over \tau}|z)\) in the denominator satisfies that \(|y|=1\). Hence, they can be approximated as follows,
\begin{align}
\theta_1(-{1\over \tau}|z)\simeq -i(y^{1\over 2}-y^{-{1\over 2}})\cdot (q^D)^{1\over 8}.
\end{align}
As a result, we obtain the following asymptotic form of the dual elliptic genus,
\begin{align}
Z_{\mathfrak{n}}\Bigr(-{1\over \tau},{m\over \tau},{\epsilon_{1,2}\over \tau}\Bigr)
=\exp\Bigr[{-2\pi i m(2p+1)-4\pi^2 (p+p^2)+\mathcal{O}(\epsilon_{1,2})\over \beta} \sum_a n_a+\mathcal{O}( \beta^0)\Bigr].
\end{align}
Then, the index (\ref{eq143_fullindex}) can be obtained from the following sum, 
\begin{align}
&Z= [Z_{U(1)}]^N    
\nonumber \\
&\times 
\sum_{\mathfrak{n} }  \exp\Bigr[ {\epsilon_1 \epsilon_2\over 2\beta }\sum_{a,b} \Omega_{a,b} n_a n_b
+\sum_a \Bigr( {(m-2\pi i p )(m-2\pi i(p+1) ) -\beta v_a+\mathcal{O}(\epsilon_{1,2})\over \beta } \Bigr) n_a
+\mathcal{O}(  \beta^0 )\Bigr].
\end{align}
By following the same procedure explained so far, we obtain the Cardy free energy at the conformal phase as follows, 
\begin{align}
Z&=\exp\Bigr[
-{N^3\over 24}{(m-2\pi ip)^2(2\pi i (p+1)- m )^2+\mathcal{O}(\beta,\epsilon_{1,2})\over \epsilon_1 \epsilon_2 \beta} \Bigr].
\end{align}
As expected, the resulting expressions are periodic under \(m\sim m+2\pi i\). This periodicity is naturally expected from the definition of the index. Note that the index counts states with \(e^{-m(Q_1-Q_2)}\). Also, \(Q_{1,2}\) are both integers for bosonic states and half-integers for fermionic states since they are spins of \(\mathbb{R}^4\) transverse to M5-branes. Hence, \(Q_1-Q_2 \in \mathbb{Z}\) which imposes a \(2\pi i\) periodicity for \(m\). See figure \ref{eq417} for the plot.

\par

Before ending this section, let us make a few comments on our chemical potential settings. In this paper, we set \(\epsilon_1>0\), \(\epsilon_2<0\), \(\beta>0\), and \(\text{Re}[m]=0\). In this setting, the expansion parameters of the index are all smaller than 1, i.e. \(e^{-\epsilon_1},e^{+\epsilon_2},e^{-\beta}<1\), which is a necessary condition for the index to be convergent. Also, the \(\mathcal{O}(1)\) imaginary part of \(m\) plays the role of obstructing \((-1)^F\) cancellation, and the resulting free energy is macroscopic. 

\par 

Technically, one can consider the generalization of the Cardy formula to \(\epsilon_{1,2}, \beta\in \mathbb{C}\). In the general complex chemical potential setting, we expect the elliptic genus sum becomes extremely difficult to evaluate with the continuum approximation. This is because that the \(Z_{\mathfrak{n}}\) can be a complex number whose phase is rapidly oscillating, and therefore, the discrete summation cannot be smoothly approximated to a continuous integral. Moreover, the elliptic genus sum can become divergent with a superfactorial growth, i.e., \(\log |Z_n|\sim n^2 \gg \log n!\) for large \(n\). In such a case, the elliptic genus expansion becomes an asymptotic expansion that should be carefully treated with the optimal truncation or resummation \cite{berry1991stokes}. Generalizing or testing our Cardy formula in such a setting seems to be technically much more challenging, and we leave it for a future problem.

\subsection{Asymptotic entropy}
\label{section24}
In this section, we compute the asymptotic entropy of (2,0) \(A\)-type theory from the Cardy free energy obtained in the previous subsection. 
\par
Let us consider the \(\mathbb{R}^4\times T^2\) index of (2,0) \(A\)-type theory defined as (\ref{eq48_index}). The index admits the fugacity expansion as follows,
\begin{align}
Z&=\text{Tr}\Bigr[(-1)^F e^{-\mathfrak{n}\cdot \mathbf{v}} e^{-\epsilon_1 J_1-\epsilon_2 J_2-\beta P } e^{-(\epsilon_+ +m)Q_1-(\epsilon_+ -m)Q_2} \Bigr]
\nonumber \\
&=\sum \Omega\cdot e^{-\mathfrak{n} \cdot \mathbf{v}}  e^{-\epsilon_1 (J_1+Q_R)-\epsilon_2 (J_2+Q_R)} e^{-\beta P} e^{-2mQ_L}
\label{eq592_index}
\end{align}
where \(Q_R\equiv {Q_1+Q_2\over 2}\) and \(Q_L\equiv {Q_1-Q_2\over 2}\). Due to the supersymmetry condition, the index can see only four charge combinations among the five possible charges \((J_1,J_2,P,Q_L,Q_R)\). Here, the expansion coefficient \(\Omega\) is the degeneracy of the microstates with \((-1)^F\) cancellation. The summation is taken over all possible charges and the self-dual string number \(\mathfrak{n}\). The microstate degeneracy can be obtained from the following contour integral,
\begin{align}
\Omega=\oint [{de^{-\epsilon_{1,2}}\over e^{-\epsilon_{1,2}}}{de^{-\beta}\over e^{-\beta}} {de^{-m}\over e^{-m}} \prod_{a=1}^{N-1}{de^{-v_a}\over e^{-v_a}} ] \cdot Z\cdot e^{ \mathfrak{n} \cdot \mathbf{v}}  e^{\epsilon_1 (J_1+Q_R)+\epsilon_2 (J_2+Q_R)} e^{\beta P} e^{2mQ_L}.
\label{eq597_degcon}
\end{align}
The contours are given by the unit circle so that they only contain the pole at the origin. The contour integral (\ref{eq597_degcon}) can be evaluated with the saddle point approximation in the Cardy limit. Let us define the entropy \(S\) as \(\log \Omega\) from the index. Then, the asymptotic entropy in the Cardy limit is given by the following expression,
\begin{align}
S\simeq \log Z+\mathfrak{n}\cdot \mathbf{v}+\epsilon_1 (J_1+Q_R)+\epsilon_2 (J_2+Q_R)+\beta P+2mQ_L
\label{eq601_extS}
\end{align}
which should be extremized with respect to the chemical potentials \(\epsilon_{1,2}\), \(\beta\), \(m\), and the tensor VEV \({v}\). Since there is a boson-fermion cancellation factor \((-1)^F\), the real part of the entropy \(\text{Re}[S]\) should be understood as the lower bound of the true degeneracy without \((-1)^F\) \cite{Choi:2018hmj, Choi:2018vbz}.
\par 
For the non-zero value of \(v_a\) satisfying \(0<v_a<{m(m-2\pi i)\over \beta}\), the Cardy free energy \(\log Z\) was computed in (\ref{eq367_freeenergy}), and it is given as follows,
\begin{align}
\log Z\simeq -{N\over 24}{m^2(2\pi i-m)^2\over \epsilon_1 \epsilon_2 \beta}
-\sum_{a,b}^{N-1} \Omega_{a,b}^{-1}{ [m(m-2\pi i)-\beta v_a][m(m-2\pi i)-\beta v_b]\over 2\epsilon_1 \epsilon_2 \beta}.
\label{eq488_freeenergy}
\end{align}
In section \ref{section2}, we derived the free energy (\ref{eq488_freeenergy}) on a special hypersurface of the space of complex chemical potentials given by (\ref{eq278_chemset}) and (\ref{eq238_rem}). Here, we will assume that (\ref{eq488_freeenergy}) holomorphically extends to complex value of the chemical potentials. Then, the extremization of (\ref{eq601_extS}) gives the following saddle point equations,
 \begin{align}
&{\partial S\over \partial \epsilon_1}=0,&
&{\partial S\over \partial \epsilon_2}=0,&
&{\partial S\over \partial \beta}=0,&
&{\partial S\over \partial m}=0, &
&{\partial S\over \partial v_a}=0.
 \end{align}
 After solving the above saddle point equations, the charges and the entropy are given as follows,
 \begin{align}
& J_1+Q_R\simeq -{N^3\over 24}{m^2(2\pi i-m)^2\over \epsilon_1^2 \epsilon_2 \beta},&
 &J_2+Q_R\simeq -{N^3\over 24}{m^2(2\pi i-m)^2\over \epsilon_1\epsilon_2^2 \beta},&
 \nonumber \\
 &Q_L\simeq {N^3\over 12}{m(m-\pi i)(m-2\pi i)\over \epsilon_1 \epsilon_2 \beta},  &
& P\simeq -{N^3\over 24}{m^2(2\pi i-m)^2\over \epsilon_1 \epsilon_2 \beta^2},&
\nonumber \\
& n_a\simeq -{m(m-2\pi i)\over 2\epsilon_1 \epsilon_2}a(N-a),&
& S\simeq -{\pi^2  N^3\over 6}{m(m-2\pi i)\over \epsilon_1 \epsilon_2 \beta}+2\pi i Q_L
\label{eq626_quntities}
 \end{align}
 where \(\simeq\) means the the above expressions hold at the leading order in the Cardy limit \(|\epsilon_{1,2}|,|\beta|\ll 1\). We set the other chemical potentials to be \(|m|,|v|\sim \mathcal{O}(1)\). Here, note the chemical potentials cannot be generic complex numbers, due to the reality condition of the charges. Indeed, all charges remain to be real in our chemical potential setting (\ref{eq278_chemset}) if we restrict \(\text{Im}[m]=\pi \). Also, the reason for our choice of the opposite signs of \(\epsilon_{1,2}\) becomes transparent, since the KK momentum \(P\) should be not only real but also be positive.
 
 \par 
 
 Now, it is straightforward to obtain the entropy \(S\) as a function of charges. The result is given as follows,
 \begin{align}
 S\simeq 2\pi \sqrt{\sqrt{ -{2\over 3} N^3 (J_1+Q_R)(J_2+Q_R)P }-Q_L^2 }+2\pi i Q_L.
 \label{eq511_entropy}
 \end{align}
In our chemical potential setting (\ref{eq278_chemset}), one can check that \(J_1+Q_R>0\) and \(P>0\), while \(J_2+Q_R<0\). This is a natural consequence since it is necessary for the index (\ref{eq592_index}) to be convergent. Therefore, the real part and the imaginary part of the entropy can be separated as follows,
\begin{align}
\text{Re}[S]\simeq 2\pi \sqrt{\sqrt{ \Bigr| {2\over 3} N^3 (J_1+Q_R)(J_2+Q_R)P }\Bigr|-Q_L^2 },\quad 
\text{Im}[S]\simeq 2\pi Q_L.
\label{eq497_ent_for}
\end{align}
One can observe that the real part of the entropy grows as the momentum-like charges \(J_{1,2}+Q_R\) and \(P\) grows. However, it decreases if the flavor-like charge \(Q_L\)increases. Recall that \(Q_L\) is the difference between the two electric charges \(Q_{1,2}\), which are the rotation of the orthogonal two planes in the transverse space of the M5-branes. Then, the entropy formula (\ref{eq497_ent_for}) tells us that if the other charges \(J_{1,2}\), \(P\), and \(Q_1+Q_2\) are fixed, the entropy is maximized at the symmetric charge configuration \(Q_1=Q_2\).

\section{Derivation from S-duality kernel}
\label{section_DSK}
In this section, we introduce the S-duality kernel of 6d (2,0) theory on \(\mathbb{R}^4\times T^2\). Then, we re-derive the results in section \ref{section2} using the S-duality kernel method. We mainly follow the formulation established in \cite{Kim:2017zyo}, and we will show that there is a noble saddle point in the S-duality kernel integral, which was not captured so far. The final results agree with those in section 2 obtained from the elliptic genera summation.
\par 
Similar to the 2d elliptic genus on \(T^2\), the 6d index on \(\mathbb{R}^4\times T^2\) also has an \(SL(2,\mathbb{Z})\) modular property. However, the \(SL(2,\mathbb{Z})\) property of the 6d index is more complicated than that of the elliptic genus. Especially, the S-duality of the \(\mathbb{R}^4\times T^2\) index is given by the convolution over the `S-duality kernel' as follows \cite{Billo:2013jba, Kim:2017zyo},
\begin{align}
Z(\tau,z,\mathbf{v})\sim \int [d\mathbf{v}^D]\cdot K(\tau,z,\mathbf{v}-\mathbf{v}^D) \cdot  Z(-{1\over \tau},{z\over \tau},\mathbf{v}^D)
\label{eq523_skernel}
\end{align}
where \(z\) collectively denotes the chemical potentials, and \(\mathbf{v}^{(D)}\) are the (dual) tensor VEVs that should be integrated out. Here, \(K\) is known as the `S-duality kernel,' and it takes the form of the Gaussian heat kernel in the prepotential limit. The S-duality relation (\ref{eq523_skernel}) is especially useful when investigating the behavior of \(Z\) at \(\tau\to i\cdot 0^+\) since the integrand can be approximated as the simple perturbative partition function in (\ref{eq170_pepert}).

\par

The S-duality kernel was first derived from the modular anomaly equation in 4d \cite{Billo:2013jba} and 6d \cite{Kim:2017zyo}. Here, we shall present a simple derivation of the S-duality kernel relation (\ref{eq523_skernel}) from the elliptic genus. Let us consider the elliptic genus expansion of the 6d index. After performing the S-dual transform of the elliptic genus, the 6d index (\ref{eq143_fullindex}) can be written as follows,
\begin{align}
Z
&=[Z_{U(1)}]^N\sum_{\mathfrak{n} }  Z_{\mathfrak{n}}e^{-\mathfrak{n}\cdot \mathbf{v}}
\nonumber \\
&=[Z_{U(1)}]^N\sum_{\mathfrak{n} }  \exp\Bigr[ {\epsilon_1 \epsilon_2\over 2\beta }\sum_{a,b} \Omega_{a,b} n_a n_b
+\sum_a ( {m^2-\epsilon_+^2\over \beta }-v_a) n_a  \Bigr]
\cdot Z_{\mathfrak{n}} \Bigr(-{1\over \tau},{m\over \tau},{\epsilon_{1,2}\over \tau}\Bigr).
\label{eq480_fulin}
\end{align}
The dual elliptic genus \(Z_{\mathfrak{n}}(-{1\over \tau},{m\over \tau},{\epsilon_{1,2}\over \tau})\) is the coefficient of \(e^{-\mathfrak{n}\cdot \mathbf{v}^D}\) in the string fugacity expansion of the dual index \(Z^D\equiv Z({1\over \tau},{m\over \tau},{\epsilon_{1,2}\over \tau},\mathbf{v}^D)\). Then, one can extract the dual elliptic genus from the dual index using the following contour integral,
\begin{align}
Z_{\mathfrak{n}}\Bigr(-{1\over \tau},{m\over \tau},{\epsilon_{1,2}\over \tau}\Bigr) ={1\over [Z_{U(1)}^D]^N} \oint_{\mathcal{C}}\prod_{a=1}^{N-1}{d(e^{-v^D_a})\over e^{-v^D_a} }\cdot e^{\mathfrak{n}\cdot \mathbf{v}^D} Z^D.
\label{eq485_coninte}
\end{align}
Here, the integral variable is the dual string fugacity \(e^{-v_a^D}\), and its contour \(\mathcal{C}\) should be taken carefully so that it only encircles the pole at the origin \(e^{-v_a^D}=0\). Inserting (\ref{eq485_coninte}) to (\ref{eq480_fulin}), we obtain the following contour integral expression,
\begin{align}
Z={[Z_{U(1)}]^N\over [Z_{U(1)}^D]^N} \oint_{\mathcal{C}} \prod_{a=1}^{N-1} {d(e^{-v_a^D})\over e^{-v_a^D}}\cdot K(\tau,m,\epsilon_{1,2},\mathbf{v}-\mathbf{v}^D)\cdot Z^D.
\label{eq491_kerint}
\end{align}
The above contour integral transforms \(Z^D\) to \(Z\). Here, \(K(\tau,m,\epsilon_{1,2},\mathbf{v}-\mathbf{v}^D)\) is the S-duality kernel which is given as follows,
\begin{align}
K(\tau,m,\epsilon_{1,2},\mathbf{x})=\sum_{\mathfrak{n}}\exp\Bigr[ {\epsilon_1 \epsilon_2\over 2\beta }\sum_{a,b} \Omega_{a,b} n_a n_b
+\sum_a ( {m^2-\epsilon_+^2\over \beta }-x_a) n_a  \Bigr].
\end{align}
In our chemical potential setting (\ref{eq278_chemset}), the \(\mathfrak{n}\) summation is convergent, and the S-duality kernel is well-defined. Then, using the continuum approximation of \(n\), we obtain the following approximate form of the S-duality kernel in the Cardy limit,
\begin{align}
K(\tau,m,\epsilon_{1,2},\mathbf{x})= \exp\Bigr[-{1 \over 2\epsilon_1 \epsilon_2 \beta}\Bigr(\sum_{a,b}\Omega^{-1}_{a,b}
( \beta x_a-m^2  )(\beta x_b-m^2 )
+\mathcal{O}(\beta,\epsilon_{1,2})\Bigr) \Bigr]
\end{align}
where we assumed that \(\text{Re}[\beta x_a]\sim\mathcal{O}(1)\) and \(\text{Im}[x_a] \in (-\pi,\pi )\). Note that we can always set \(\text{Im}[x_a]\in(-\pi,\pi)\) since the S-duality kernel is periodic under \(x\sim x+2\pi i\). As we can see, the leading term of the S-duality kernel takes the form of the Gaussian which is equivalent to the heat kernel solution of the modular anomaly equation of the \(\mathbb{R}^4\times T^2\) prepotential \cite{Kim:2017zyo}. 
\par
We shall evaluate the S-duality kernel integral (\ref{eq491_kerint}) using the saddle point approximation. In the Cardy limit where \(\beta\ll 1\), the dual index \(Z^D\) has a exponentially small instanton fugacity, i.e. \(q^D=e^{2\pi i \tau^D}\ll 1\). Therefore, one might guess that the dual instanton partition function can be ignored. However, we should be careful about estimating the dual instanton partition function because there are other fugacities which can give growth, or suppression factors comparable to \(q^D\). First, the dual mass fugacity \(e^{-m^D}\) can have the same order with the instanton fugacity since \(m^D={m\over \tau}=\mathcal{O}(\tau^D)\). Second, the dual string fugacity \(e^{-v^D}\) should also be considered because \(v^D\) can have a large real part at the saddle point of the S-duality kernel integral (\ref{eq491_kerint}). As we shall see, it turns out that the saddle point satisfies \(|e^{-v_a^D}|\leq |e^{-m^D}|\ll 1\). With this assumption, the \(k\)-instanton partition function has the following order,
\begin{align}
Z^D_k \cdot (q^D)^k = \mathcal{O}\Bigr( (q^D e^{m^D})^k \Bigr)=\mathcal{O}\Bigr( \exp\Bigr[-k{4\pi^2+2\pi im\over \beta}\Bigr] \Bigr).
\label{eq509_inst_supp}
\end{align}
Therefore, the dual instanton partition function can be ignored only in the canonical chamber \(0<\text{Im}[m]<2\pi \). This is consistent with our observation in (\ref{eq389_outsi}) and (\ref{eq395_dfkbdm}) where the \(q^D\) corrections in the dual elliptic genus becomes negligible only inside the canonical chamber. We will see that our assumption \(|e^{-v_a^D}|\leq |e^{-m^D}|\ll 1\) indeed holds for the saddle point value of \(v^D\).
\par 
Then, we can approximate the dual index as the perturbative index only. We further take the prepotential limit \(\epsilon_{1,2}^D\to 0\). It can be achieved by setting \(|\epsilon_{1,2}|\ll |\beta|\), and the leading free energy in the Cardy limit is not affected by the relative scaling between \(\epsilon_{1,2}\) and \(\beta\). In the prepotential limit, \(Z^D\) can be written as follows,
\begin{align}
Z^D&=[Z_{U(1)}^D]^N\text{PE}\Bigr[{2\sinh{m\pm\epsilon_+\over 2\tau}\over 2\sinh{\epsilon_{1,2}\over 2\tau}}\sum_{\alpha\in\Psi^+}e^{-\alpha\cdot \mathbf{v}^D} \Bigr]\cdot \Bigr(1+\mathcal{O}(q^D)\Bigr)
\nonumber \\
&=[Z_{U(1)}^D]^N\exp\sum_{\alpha\in\Psi^+}\Bigr[ {\tau^2\over \epsilon_1 \epsilon_2} \Bigr( \text{Li}_3(e^{-\alpha\cdot \mathbf{v}^D\pm {m\over \tau}})-2\text{Li}_3(e^{-\alpha\cdot \mathbf{v}^D}) \Bigr)
\nonumber \\
&-{1\over 24}\Bigr\{ ({\epsilon_1\over \epsilon_2}+{\epsilon_2\over \epsilon_1} )\Bigr(4\text{Li}_1(e^{-\alpha\cdot \mathbf{v}^D})+\text{Li}_1(e^{-\alpha\cdot \mathbf{v}^D\pm {m\over \tau}}) \Bigr) +12\text{Li}_1(e^{-\alpha\cdot \mathbf{v}^D}) \Bigr\}
+\mathcal{O}(\epsilon_{1,2}^2)  \Bigr]  .
\label{eq520_prepot}
\end{align}
We denoted the suppressed instanton contribution as \(\mathcal{O}(q^D)\). The second line of (\ref{eq520_prepot}) is the prepotential, which gives the leading free energy in the thermodynamic limit \cite{Kim:2017zyo}. Here, we expanded the free energy beyond the leading prepotential for the reasons that will be explained now.
\par 
The subleading terms in the third line of (\ref{eq520_prepot}) are not important when considering the free energy itself, but they play an important role when performing the saddle point analysis of the S-duality kernel integral. Note that \(\text{Li}_1(x)=-\log(1-x)\), and one can rewrite (\ref{eq520_prepot}) as follows,
\begin{align}
Z^D&=[Z^D_{U(1)}]^N
\prod_{\alpha\in\Psi^+}
\Bigr(1-e^{-\alpha\cdot \mathbf{v}^D\pm {m\over \tau} } \Bigr)^{\epsilon_1^2+\epsilon_2^2\over 24\epsilon_1 \epsilon_2}
\Bigr(1-e^{-\alpha\cdot \mathbf{v}^D} \Bigr)^{\epsilon_1^2+3\epsilon_1 \epsilon_2+\epsilon_2^2\over 6\epsilon_1 \epsilon_2}
\nonumber \\
&\times \exp\Bigr[\sum_{\alpha\in\Psi^+} {\tau^2\over \epsilon_1 \epsilon_2}\Bigr(\text{Li}_3(e^{-\alpha\cdot \mathbf{v}^D \pm{m\over \tau} })
-2\text{Li}_3(e^{-\alpha\cdot \mathbf{v}^D}) \Bigr)+\mathcal{O}(\epsilon_{1,2}^2)\Bigr].
\label{eq526_prepoten}
\end{align}
Note that the dual index has branch-point singularities at \(e^{-\alpha\cdot \mathbf{v}^D}= e^{\pm{m/ \tau}}\) and \(e^{-\alpha\cdot \mathbf{v}^D}=1\). Those singularities cannot be captured if we investigate the prepotential only, and the existence of the branch-points critically affects the saddle point structure of the integral.
\par
\begin{figure}[t!]
\centering
\begin{subfigure}[a]{.49\textwidth}
\includegraphics[width=0.9\columnwidth]{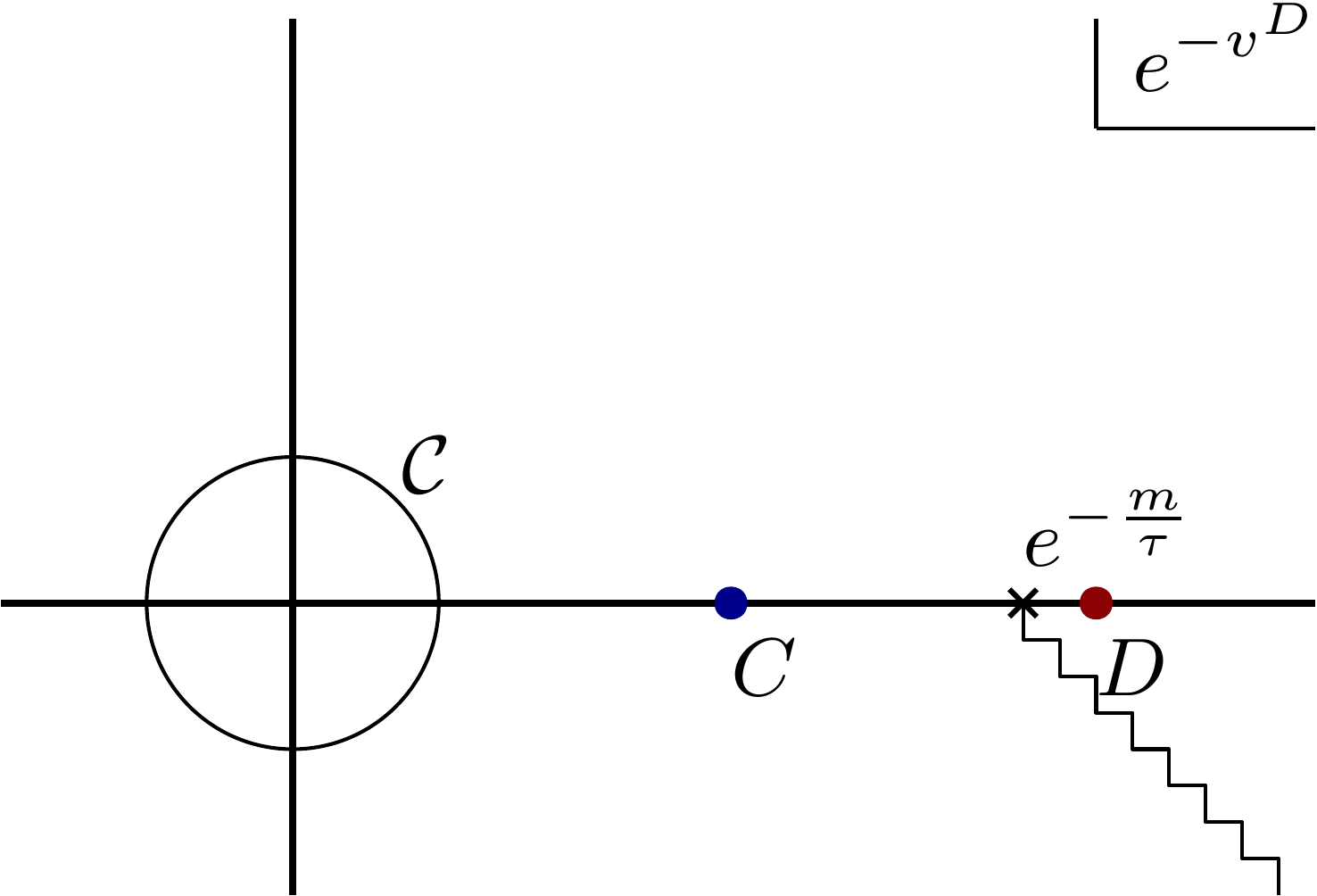}
\caption{\(v>{m(m-2\pi i)\over \beta}\)}
\end{subfigure}
\begin{subfigure}[a]{.49\textwidth}
\includegraphics[width=0.9\columnwidth]{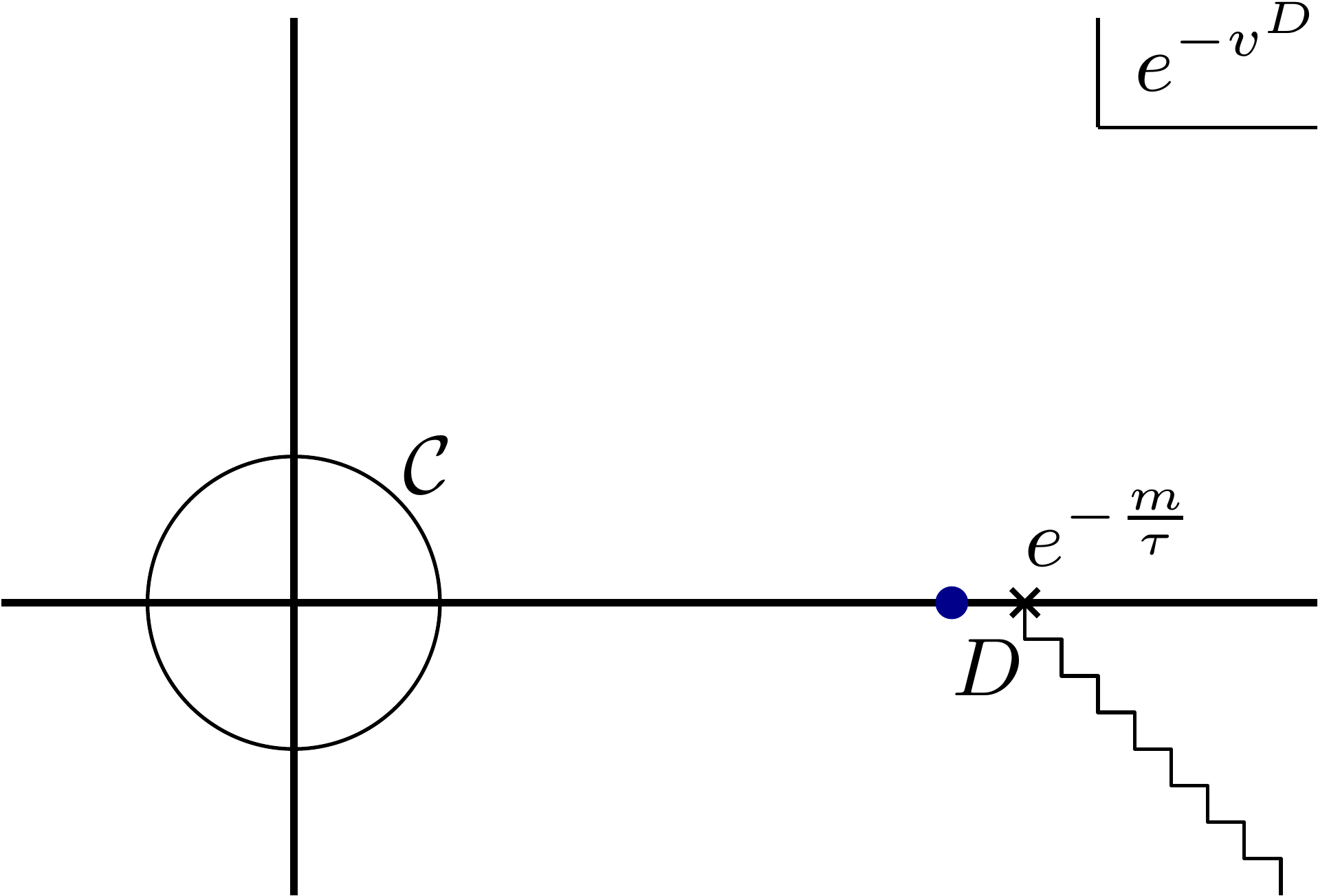}
\caption{\({m(m-2\pi i)\over \beta}>v>0\)}
\end{subfigure}
\caption{Contour \(\mathcal{C}\) of the S-duality kernel integral. The cross denotes the branch point at \(e^{-v^D}=e^{-m/\tau}\), and the wavy line denotes the branch cut whose direction is set not to cross the saddle points. The points \(C\) and \(D\) denote the confining and deconfining saddle points, respectively. The blue saddle point is the accessible saddle point, and the red saddle point is the inaccessible saddle point.}
\label{fig541}
\end{figure}

 
The saddle point approximation requires the deformation of the original contour to the steepest descent contour, which passes the saddle point. Therefore, the original contour should be concretely defined to perform the saddle point analysis. Now, let us explain how the contour \(\mathcal{C}\) in the S-duality kernel integral (\ref{eq491_kerint}) should be set. As we mentioned, the contour \(\mathcal{C}\) should only encircles the pole at the origin \(e^{-v_a}=0\). Also, the contour should not pass through the branch-cut for the integral to be well-defined. The integrand \(Z^D\) has multiple branch-cuts which start from the branch-point \(e^{-\alpha\cdot \mathbf{v}^D}=e^{\pm m/\tau},1\) and they extend to \(\text{Re}[\alpha\cdot \mathbf{v}^D]\to \infty\). Then, we can set a proper contour \(\mathcal{C}\) as a small circle \(|e^{-v_a|}=C\) such that \(C<e^{-m/\tau}\). The structure of the contour is illustrated in figure \ref{fig541}.
\par 
Now, we deform the contour \(\mathcal{C}\) to the steepest descent contour that passes through saddle points. The saddle point equation is given as follows,
\begin{align}
&0={\partial\over \partial v_a^D}\Bigr[\log K(\tau,m,\epsilon,v-v^D)+\log Z^D\Bigr]
\nonumber \\
&={1\over \epsilon_1 \epsilon_2}\sum_b \Omega^{-1}_{a,b}\Bigr( \beta(v_a-v_a^D)-m^2 \Bigr)
+\sum_{\alpha\in \Psi^+}{\partial(\alpha\cdot \mathbf{v}^D)\over \partial v_a^D}\Bigr[ -{\tau^2\over \epsilon_1 \epsilon_2}\Bigr(\text{Li}_2(e^{-\alpha\cdot \mathbf{v}^D\pm{m\over \tau} })-2\text{Li}_2(e^{-\alpha\cdot \mathbf{v}^D })\Bigr)
\nonumber \\
&+{\epsilon_1^2+\epsilon_2^2\over 24\epsilon_1 \epsilon_2}{e^{-\alpha\cdot \mathbf{v}^D \pm {m\over \tau} } \over 1-e^{-\alpha\cdot \mathbf{v}^D \pm {m\over \tau} } }
+{\epsilon_1^2+3\epsilon_1 \epsilon_2+\epsilon_2^2\over 6\epsilon_1 \epsilon_2}{e^{-\alpha\cdot \mathbf{v}^D  } \over 1-e^{-\alpha\cdot \mathbf{v}^D   } } \Bigr]+\mathcal{O}(\epsilon_{1,2}^2).
\label{eq545_spe}
\end{align}
The conventional saddle points can be obtained if we only consider the leading \(\mathcal{O}({1\over \epsilon_1 \epsilon_2})\) terms in the first line of (\ref{eq545_spe}). For simplicity, let us assume that all \(v_a\)'s are the same so that the saddle point of \(v_a^D\)'s are the same for \(1\leq a \leq N-1\). Then, we can denote \(v_a=v\) and \(v_a^D=v^D\). Note that all polylogarithms can be ignored when \(v^D>{m\over \tau}\). In this regime, the saddle point equation becomes the linear equation whose solution is given as follows,
\begin{align}
\text{confining saddle: }v^D=v-{m^2\over \beta}\quad \text{if}\quad v>{m(m-2\pi i)\over \beta}.
\label{eq559_confined}
\end{align}
The above solution is valid when \(v>{m(m-2\pi i)\over \beta}\) since we assumed that \(v^D>{m\over \tau}\). For the reasons to be clear shortly, we will label the above saddle point as a confining saddle point. The confining saddle point satisfies \(e^{-v^D}<e^{-m/\tau}\) and the instanton suppression condition (\ref{eq509_inst_supp}) is satisfied. 
\par 
When \(v<{m(m-2\pi i)\over \beta}\), the effect of the polylogarithm functions cannot be ignored, and we should find other saddle points. The saddle point analysis in this setting was first studied in \cite{Kim:2017zyo}. In this paper, we report the presence of the saddle point other than the one found in \cite{Kim:2017zyo}. Our new saddle point corresponds to the deconfining 6d free energy proportional to \(N^3\).
\par 
For the careful analysis of the saddle point approximation, one should take the effect of the second line in (\ref{eq545_spe}) into account.  Although it naively seems subleading in the Cardy limit, it can provide a comparable contribution of order \(\mathcal{O}({1\over \epsilon_1 \epsilon_2})\) when \(v^D\) is sufficiently close to the singularity. Such saddle points are known as the saddle point near singularity \cite{temme2015asymptotic, lee2013modified}. Among the many possible singularities, the one that is closest to the contour \(\mathcal{C}\) is located at \(e^{-v^D}=e^{-m/\tau}\). By solving the saddle point equation (\ref{eq545_spe}) near \(v^D={m\over \tau}\), one can find the following saddle point solution,
 \begin{align}
&\text{deconfining saddle: } v^D={m\over \tau}+{1\over 12 a(N-a) }{\epsilon_1^2+\epsilon_2^2\over m(m-2\pi i)-\beta v}+O(\epsilon_{1,2}^4) .
 \end{align}
We label the above saddle point as the deconfining saddle. The deconfining saddle point solves the saddle point equation regardless of whether \(v\) is smaller then \({m(m-2\pi i)\over \beta}\) or not. However, the deformation of the original contour to the steepest descent contour passing the deconfining saddle is possible only when \(0<v<{m(m-2\pi i)\over \beta}\). Note that when \(v>{m(m-2\pi i)\over \beta}\), the deconfining saddle lies outside of the singularity, i.e. \(e^{-v^D}>e^{-m/\tau}\). In this case, the contour deformation to pass the deconfining saddle is forbidden by the branch-cut starting from \(e^{-v^D}=e^{-m/\tau}\). See figure \ref{fig541} for the illustration. Such saddle point is known as an `inaccessible saddle point' \cite{oughstun2009analysis}. On the contrary, when \(0<v<{m(m-2\pi i )\over \beta}\), the deconfining saddle lies inside of the singularity, i.e. \(e^{-v^D}<e^{-m/\tau}\). Then, the contour can be deformed to pass the deconfining saddle, and it can contribute to the S-duality kernel integral\footnote{The saddle point found in \cite{Kim:2017zyo} satisfies the saddle point equation (\ref{eq545_spe}) at the leading order. However, it lies outside of the singularity, and therefore, it is inaccessible. }.
\par 
To summarize, the accessible saddle point is the confining saddle point when \(v>{m(m-2\pi i)\over \beta}\) and the deconfining saddle point when \({m(m-2\pi i)\over \beta}>v>0\). The saddle point structure is illustrated in figure \ref{fig541}.
 \par 
Now, let us compute the free energy from the saddle points obtained so far. The leading free energy can be obtained by inserting the saddle point value to the integrand in (\ref{eq491_kerint}). Therefore, the resulting free energy becomes
\begin{align}
&\log Z
\nonumber \\
=&[\log Z]_\text{C}\simeq -{N\over 24}{m^2(2\pi i-m)^2\over \epsilon_1 \epsilon_2 \beta},
& \Bigr({m(m-2\pi i)\over\beta}<v\Bigr)&
\nonumber \\
=&[\log Z]_\text{D}
\simeq  -{N\over 24}{m^2(2\pi i-m)^2\over \epsilon_1 \epsilon_2 \beta}
-{N^3-N\over 24}{[m(m-2\pi i)-\beta v]^2 \over \epsilon_1 \epsilon_2 \beta},
& \Bigr(0<v<{m(m-2\pi i)\over\beta}\Bigr)&
\end{align}
where the subscript \([\log Z]_\text{C/D}\) stands for confining/deconfining saddle point. As can be seen, the confining saddle point gives the Abelian free energy, while the deconfining saddle point gives the non-Abelian contribution. As we can see, the result is the same as the phenomena observed in section \ref{section22}.

\section{E-string theory of rank \(N\)}
\label{section3}
\subsection{\(\mathbb{R}^4\times T^2\) index}
\begin{figure}
\centering
\includegraphics[width=0.45\columnwidth]{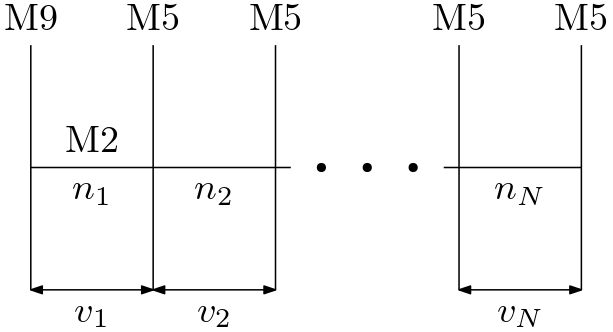}
\caption{M9-M5 brane set-up for the rank \(N\) E-string theory}
\label{fig721_Estring}
\end{figure}

A 6d (1,0) rank \(N\) E-string theory is a worldvolume theory of \(N\) parallel M5-branes probing M9-brane \cite{Horava:1995qa, Klemm:1996hh}. The transverse space of the M5-brane hosts \(SO(4)\simeq SU(2)_{R} \times SU(2)_{L}\) symmetry where \(SU(2)_{R}\) is the R-symmetry of the (1,0) supersymmetry and \(SU(2)_{L}\) is a flavor symmetry. Also, there is an \(E_8\) global symmetry from the M9-brane. In the tensor branch, multiple M2-branes are suspended between two M5-branes, or between the M5-brane and the M9-brane. The latter becomes the self-dual string in 6d, which is called the E-string \cite{Minahan:1998vr}.
\begin{align}
Z=\text{Tr}\Bigr[ (-1)^F e^{-\beta k} e^{-\mathfrak{n}\cdot \mathbf{v} } e^{-\epsilon_1 J_1-\epsilon_2 J_2} e^{-2\epsilon_+ Q_R-2m Q_L }\prod_{l=1}^8 e^{-m_l F_l}\Bigr]
\label{eq216_Estring}
\end{align}
where \(k\) is the KK momentum, \(J_{1,2}\) are angular momenta on \(\mathbb{R}^4\), and \(Q_{L,R}\)'s are Cartan charges of \(SU(2)_{L,R}\) respectively. There is also the \(E_8\) global symmetry whose Cartan charges are denoted as \(F_l\)'s. Lastly, \(\mathbf{v}=(v_1,...,v_N)\) is the tensor VEV and \(\mathfrak{n}=(n_1,...,n_N)\) is the charge of the self-dual string. Here, \(v_1\) parameterizes the distance between the M9-brane and the first M5-brane, and \(v_{a>1}\) parameterizes  the distance between the \(a-1\)'th and \(a\)'th M5-branes. Similarly, \(n_1\) is the number of the M2-branes between M9 and M5, and \(n_{a>1}\) is the number of M2-branes between the \(a-1\)'th and \(a\)'th M5-branes. The brane set-up is illustrated in figure \ref{fig721_Estring}.

\par 
The \(\mathbb{R}^4\times T^2\) index can be expanded with elliptic genera of the E-strings as follows,
\begin{align}
Z(v,\tau,\epsilon_{1,2},m,m_l)=[Z_{U(1)}(\tau,\epsilon_{1,2},m)]^N\cdot \sum_{\mathfrak{n}} Z_{\mathfrak{n}} (\tau,\epsilon_{1,2},m,m_l) e^{-\mathfrak{n}\cdot \mathbf{v} }.
\label{eq220_gci}
\end{align}
Here, \(Z_{\mathfrak{n}}\) is the elliptic genus of the self-dual strings with charge \(\mathfrak{n}\). The elliptic genus can be computed from the 2d (0,4) gauge theories with gauge group \(U(n_1)\times U(n_2)\times ... \times U(n_N)\) \cite{Kim:2014dza, Gadde:2015tra,Kim:2015fxa}. The fully refined elliptic genus can be computed from the following contour integral,
\begin{align}
&Z_{\mathfrak{n}}=\oint [\prod_{a=1}^{N}{1\over n_a!}\prod_{i=1}^{n_a}d\phi^a_i] 
\Bigr( {\eta(\tau)^2 \theta(2\epsilon_+)\over \theta(\epsilon_1)\theta(\epsilon_2)}\Bigr)^{\sum_{a=1}^N n_a}
\nonumber \\
&\times \prod_{i\neq j}^{n_1} {\theta(\phi^1_{ij})\theta(\phi^1_{ij}+2\epsilon_+)\over \theta(\phi^1_{ij}+\epsilon_{1,2}) }
\prod_{i>j}^{n_1}
{\theta( \epsilon_- \pm (\phi_i+\phi_j+M))\over \theta(-\epsilon_+ \pm (\phi_i+\phi_j+M) }
 \prod_{i=1}^{n_1} 
 { \theta(\phi_i^1 -m) \theta(\phi_i^1+M)  \prod_{l=1}^8 \theta(\phi^1_i-\hat{m}_l)
 \over \theta(-\epsilon_+ \pm(2\phi_i+M)) \theta(\epsilon_+ \pm \phi^1_i )}
\nonumber \\
&\times\prod_{a=2}^N \Bigr[ \prod_{i \neq j}^{n_a}{\theta(\phi^a_{ij})\theta(\phi^a_{ij}+2\epsilon_+) \over \theta(\phi^a_{ij} +\epsilon_{1,2}) }
\prod_{i=1}^{n_a} {
\theta(\phi_i-am) \theta(\phi+(a-2)m)
\over \theta(\epsilon_+ \pm (\phi^a_i-(a-1)m ) ) }
\prod_{i=1}^{n_a} \prod_{j=1}^{n_{a-1}} {\theta(\epsilon_- \pm (\phi^a_i -\phi^{a-1}_j)) \over \theta(-\epsilon_+ \pm (\phi^a_i -\phi^{a-1}_j)) }\Bigr]
\label{eq673_integrand}
\end{align}
where \(\phi_i^a\) is the \(i\)'th gauge holonomy of \(U(n_a)\) gauge node. The chemical potential mapping is given by \(M=-3m-\sum_l m_l\), and \(\hat{m}_l=m+m_l+{1\over 4}\sum_k m_k\). Here, the function \(\theta(z)\) is defined as \(\theta(z)\equiv (i\eta(\tau))^{-1}{\theta_1(\tau|{z\over -2\pi i})}\), and we used a shorthand notation that \(\theta(x\pm y)=\theta(x+y)\theta(x-y)\) and \(\theta(x+y_{1,2})=\theta(x+y_1)\theta(x+y_2)\). The pole prescription is given by the Jeffrey-Kirwan residue \cite{Benini:2013xpa}.
\par 
The S-duality property of the elliptic genus can be obtained from 2d chiral anomaly on the E-string, which can be obtained from the 6d anomaly polynomial of 6d E-string theory. The anomaly 8-form \(I_8\) of the 6d E-string theory is given as follows \cite{Ohmori:2014kda},
\begin{align}
I_8&=I_\text{1-loop}+{1\over 2} \sum_{a,b=1}^N \Omega^{-1}_{a,b} I_4^a I_4^b
\nonumber \\
I_\text{1-loop}&={N\over 48}\Bigr[ p_2(N)-p_2(T)+{1\over 4}\Bigr(p_1(T)-p_1(N) \Bigr)^2 \Bigr]
\nonumber \\
I_4^{a=1}&=-{1\over 4}\Bigr( {1\over (2\pi)^2}\text{Tr}[\mathbf{F}_{E_8}^2]+p_1(T)+p_1(N) \Bigr)-{1\over 2}\chi_4(N)
\nonumber \\
I_4^{a>1}&= \chi_4(N).
\label{eq660_anom_Estring}
\end{align}
Here, \(N\) stands for \(SO(4)_N=SU(2)_R\times SU(2)_L\) normal bundle, and \(T\) stands for \(SO(4)_T=SU(2)_r\times SU(2)_l\) tangent bundle. \(p_{1,2}\) are Pontryagin classes and \(\chi_4\) is the Euler character. Also, \(\Omega_{a,b}^{-1}\) denotes the \((a,b)\) component of \(\Omega^{-1}\). The \(N\times N\) intersection matrix \(\Omega\) is defined as follows,
\begin{align}
\Omega=\begin{pmatrix}
1 & -1 & 0 & ... & 0 & 0 \\
-1& 2 & -1 & ... & 0 & 0 \\
0 & -1 & 2 & ... & 0 & 0 \\
... \\
0 & 0 & 0 & ... & 2 & -1 \\
0 & 0 & 0 & ... & -1 & 2 \\
\end{pmatrix}.
\end{align}
Note that \(\Omega\) is not a Cartan matrix of any Lie algebra\footnote{Although E-string theory on \(S^1\) becomes 5d \(Sp(N)\) theory with 8 fundamentals, the intersection matrix \(\Omega\) differs from the \(Sp(N)\) Cartan matrix due to \(E_8\) flavor symmetry of M9-brane.}. For the 2d CFT on the self-dual strings, its anomaly can be computed by matching with the anomaly inflow from 6d. Then, the 2d anomaly on \(\mathfrak{n}\) E-string is given by \cite{Kim:2016foj},
\begin{align}
I_4^{2d}=&-{1\over 2}\chi_4(T)  \sum_{a,b} \Omega_{a,b} n_a n_b -\sum_a I_4^a  n_a
\nonumber \\
=&-{1\over 2}\Bigr( c_2(l)-c_2(r)  \Bigr) \sum_{a,b}\Omega_{a,b} n_a n_b+\Bigr(c_2(L)-c_2(R)\Bigr) \sum_{a=2}^N n_a
\nonumber \\
&+\Bigr[ {1\over 4}\Bigr( {1\over (2\pi)^2}\text{Tr}[\mathbf{F}^2_{E_8}]-2[c_2(l)+c_2(r)] -2[c_2(L)+c_2(R)] \Bigr)
+{1\over 2}\Bigr(c_2(L)-c_2(R) \Bigr) \Bigr]n_1
\end{align}
where we used that \(p_1(N)=-2[c_2(L)+c_2(R)]\), \(\chi_4(N)=c_2(L)-c_2(R)\), and similar relations for the tangent bundle. The S-dual anomaly of the elliptic genus is obtained by replacing the characteristic classes in \(I_4^{2d}\) to the corresponding chemical potentials as follows,
\begin{align}
Z_{\mathfrak{n}}(\tau,\epsilon_{1,2},m_l)&=\exp\Bigr[{\pi i\over \tau} \mathfrak{i}_{\mathfrak{n}} \Bigr]
\cdot Z_{\mathfrak{n}}(-{1\over \tau},{\epsilon_{1,2}\over \tau},{m_l\over \tau}).
\label{eq245_sdual}
\end{align}
Here, \(\mathfrak{i}_{\mathfrak{n}}\) is the quadratic polynomial of the chemical potentials which can be obtained from the following replacement rule,
\begin{align}
\mathfrak{i}_{\mathfrak{n}}={1\over 2\pi^2}I_4^{2d}\Bigr(c_2(l)\to \epsilon_-^2,c_2(L)\to m^2,c_2(r/R)\to \epsilon_+^2, {1\over 2(2\pi)^2}\text{Tr}[\mathbf{F}_{E_8}^2]\to \sum_l m_l^2 \Bigr).
\end{align}
As a result, we can derive the S-dual property of the elliptic genus as follows,
\begin{align}
&Z_{\mathfrak{n}}(\tau,\epsilon_{1,2},m_l)
\nonumber \\
=&\varepsilon^{-6n_1} \exp\Bigr[-{1\over 4\pi i \tau}\Bigr(\epsilon_1 \epsilon_2 \sum_{a,b} \Omega_{a,b} n_a n_b
+(\sum_{l=1}^8 m_l^2-3\epsilon_+^2-\epsilon_-^2)n_1+2(m^2-\epsilon_+^2)\sum_{a=2}^N n_a \Bigr) \Bigr]
\nonumber \\
\times &Z_{\mathfrak{n}}(-{1\over \tau},{\epsilon_{1,2}\over \tau},{m_l\over \tau}).
\label{eq779_sdual}
\end{align}
The above S-duality property can be explicitly checked from the integral expression (\ref{eq673_integrand}) also. The phase factor \(\epsilon\) is a constant defined by \({\eta(-1/\tau)\over \theta_1(-1/\tau,z/\tau)}=\varepsilon e^{-{\pi i z^2\over \tau}}{\eta(\tau)\over \theta_1(\tau,z)} \) \cite{Kim:2014dza}.

\subsection{Free energy in the Cardy limit}

As we have studied in the previous section, we are interested in the free energy of the index in the Cardy limit. Hence, we shall take our chemical potentials to satisfy (\ref{eq230_highT}) and (\ref{eq278_chemset}). Similarly, we set the flavor symmetry chemical potential \(m_l\)'s to be purely imaginary. We will also take \(v\sim\mathcal{O}(\beta^{-1})\) as we have done in section \ref{section22}, where \(\beta=-2\pi i \tau\).
\par 
Using the S-duality of the E-string elliptic genus (\ref{eq779_sdual}), we rewrite the index (\ref{eq220_gci}) as follows,
\begin{align}
Z
&=[Z_{U(1)}]^N \sum_{\mathfrak{n} }  Z_\mathfrak{n} e^{-\mathfrak{n}\cdot \mathbf{v} }
\nonumber \\
&=[Z_{U(1)}]^N \sum_{\mathfrak{n} }  \exp\Bigr[{ \epsilon_1 \epsilon_2\over 2\beta} \sum_{a,b} \Omega_{a,b} n_a n_b
+{\sum m_l^2-3\epsilon_+^2-\epsilon_-^2 -2\beta v_1\over 2\beta}n_1+\sum_{a=2}^N {m^2-\epsilon_+^2-\beta v_a\over \beta}n_a \Bigr] 
  Z_{\mathfrak{n}}^D
\label{eq307_fullindex}
\end{align}
where \(Z^D_{\mathfrak{n}}=Z_{\mathfrak{n}}(-{1\over \tau},{\epsilon_{1,2}\over \tau},{m\over \tau},{m_l\over \tau})\) is the dual elliptic genus. For simplicity, we shall perform the computation in the following regime which we call a `canonical chamber,'
\begin{align}
0<\text{Im}[m]<2\pi,\quad -{\pi\over 2}<\text{Im}[m_l]<{\pi \over 2},\quad (1\leq l \leq 8).
\label{eq369_canonicalchamber}
\end{align}
Inside the canonical chamber, the dual elliptic genus has a simple asymptotic form in the Cardy limit. When \(\tau\to i\cdot 0^+\), the dual modular parameter becomes \(\tau^D\to i\cdot \infty\), and the leading behavior is determined as follows, 
\begin{align}
 Z_{\mathfrak{n}}(-{1\over \tau},{\epsilon_{1,2}\over \tau},{m\over \tau},{m_l\over \tau})=\exp\Bigr[{1\over \beta}\Bigr( ({2\pi^2 }+\mathcal{O}(\epsilon_{1,2})) n_1-({2\pi i m}+\mathcal{O}(\epsilon_{1,2}))\sum_{a=2}^N n_a\Bigr)+\mathcal{O}(\beta^0) \Bigr] 
 \label{eq315_elliptic}
\end{align}
which is explicitly checked for the elliptic genera up to \(n_1\leq 4\) for the rank 1 results in \cite{Kim:2014dza}, and up to \((n_1,n_2)\leq (1,1)\) for the rank 2 results in \cite{Kim:2015fxa}. Using the above asymptotic form, one can reorganize the index (\ref{eq307_fullindex}) as follows,
\begin{align}
Z=[Z_{U(1)}]^N \sum_{\mathfrak{n}} \exp\Bigr[ {\epsilon_1 \epsilon_2\over 2\beta}\sum_{a,b} \Omega_{a,b} n_a n_b +{1\over \beta }\sum_a (\rho_a-\beta v_a+\mathcal{O}(\epsilon_{1,2}) ) n_a +\mathcal{O}(\beta^0) \Bigr].
\label{eq723_estring_sum}
\end{align}
where we introduce a vector \(\rho_a\) s follows,
\begin{align}
\rho_1=2\pi^2+{\sum_l m_l^2 \over 2},\quad 
\rho_{a>1}=m(m-2\pi i) .
\end{align}
The string charge summation in (\ref{eq723_estring_sum}) also takes the form of the Gaussian. Now, we evaluate the elliptic genus summation with the continuum approximation. Let us define a variable \(x\) as \(x_a \equiv -\epsilon_1 \epsilon_2 n_a\) which becomes a continuous variable in the Cardy limit. Then, the summation (\ref{eq723_estring_sum}) can be approximated to the following integral,
\begin{align}
Z=[Z_{U(1)}]^N \int_0^\infty {dx\over |\epsilon_1 \epsilon_2|} \exp\Bigr[{1\over \epsilon_1 \epsilon_2 \beta}\Bigr( {1\over 2}\sum \Omega_{a,b} x_a x_b -\sum_a ({\rho}_a-\beta v_a) x_a+\mathcal{O}(\epsilon_{1,2})\Bigr)+\mathcal{O}( \beta^0)  \Bigr].
\label{eq775_inte}
\end{align}
As we have done in section \ref{section21}, we can evaluate the above integral with the saddle point approximation. Then, we obtain the following saddle point solution,
\begin{align}
x_a=\sum_b \Omega^{-1}_{a,b} ({\rho}_b-\beta v_b)
\end{align}
Note that the integral (\ref{eq775_inte}) can be approximated to the value at the above saddle point only when all \(x_a\)'s are inside the integral range \(x_a\in(0,\infty)\). Such condition is satisfied when \(v_a\ll \mathcal{O}(\beta^{-1})\), and the corresponding free energy is given as follows,
\begin{align}
\log Z=-{1\over 24}{m^2(2\pi i-m)^2\over \epsilon_1 \epsilon_2 \beta}
-{1\over 2\epsilon_1 \epsilon_2 \beta}\sum_{a,b} \Omega^{-1}_{a,b} ({\rho}_a-\beta v_a) ({\rho}_b-\beta v_b)+{ \mathcal{O}(\epsilon_{1,2}) \over \epsilon_1 \epsilon_2 \beta } .
\label{eq791_freeenergy}
\end{align}
\par
\begin{figure}
\centering
\includegraphics[width=0.5\columnwidth]{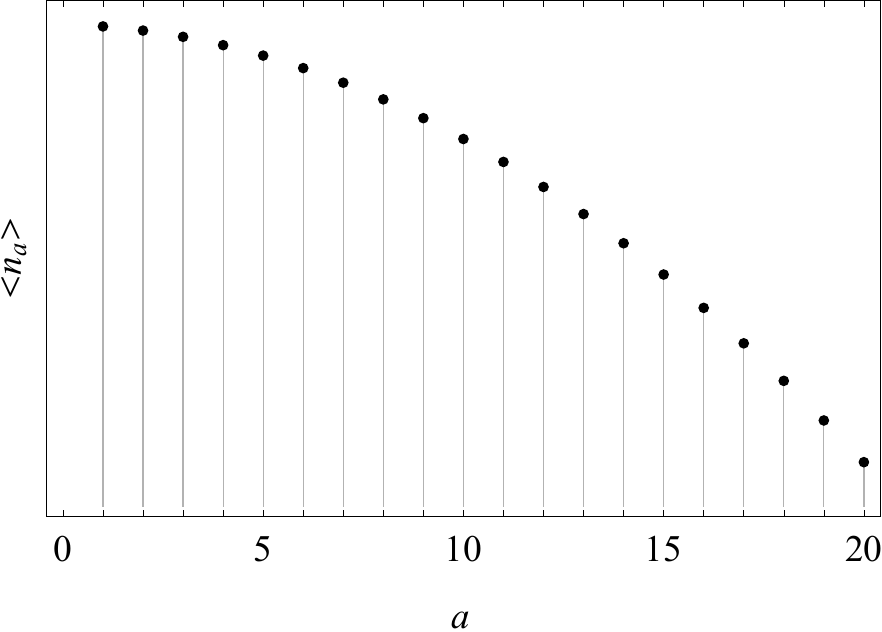}
\caption{The string condensation VEV \(\langle n_a \rangle\) for the rank \(N=20\) E-string theory. We set \(m_l=0\) and \(m=\pi i\) for the simplicity. One can see that the strings are more condensed near the M9-brane. }
\label{fig861}
\end{figure}

 
Now, let us take the limit \(\beta v_a \to 0\) to focus on the physics at the conformal phase. Then, the relevant saddle point solution is given as follows,
\begin{align}
x_a=(N+1-a)\cdot (2\pi^2+ {\sum_l m_l^2\over 2} )+  {N^2-N-2+3a-a^2\over 2}\cdot m(m-2\pi i)
\label{eq790_saddle}
\end{align}
The above saddle point is related to the charge condensation of the self-dual string by \(\langle n_a \rangle =(-\epsilon_1 \epsilon_2)^{-1} x_a\). In the case of (2,0) \(A_{N-1}\)-type theory, we found that \(\langle n_a \rangle \propto a(N-a)\) which is symmetric under \(a\to N-a\). In this case, due to the presence of the M9-brane wall, the self-dual strings are more condensed near the M9-brane. See figure \ref{fig861} for the plot. One can find two distinct combinatoric factors \(N+1-a\) and \({N^2-N-2+3a-a^2\over 2}\), which are both integers.  Since the factor \(N+1-a\) is charged under the \(E_8\) flavor symmetry, we speculate that it is related to the M2-branes suspended between M9 and M5-brane carrying KK momentum. On the other hand, \({N_2-N-2+3a-a^2\over 2}\) is related to the M2-branes suspended between two M5-branes, as discussed in section \ref{section22}. Sum over those factors are given as follows,
\begin{align}
\sum_{a=1}^N (N+1-a)={N^2+N\over 2},\quad \sum_{a=1}^{N} {N^2-N-2+3a-a^2\over 2}={N^3-N\over 3}.
\end{align}
Surprisingly, we encounter with \({N^3-N}\) factor again for the degrees on \(N\) M5-branes. 
\par 
Lastly, by evaluating the free energy (\ref{eq791_freeenergy}) at the conformal phase \(\beta v \ll 1\), we obtain the following Cardy formula,
\begin{empheq}[box=\widefbox]{align}
\log Z=&-{N\over 8}{(4\pi^2+ \sum_l m_l^2)^2\over \epsilon_1 \epsilon_2 \beta}-{N(N-1)\over 4}{(4\pi^2+\sum_l m_l^2)m(m-2\pi i)\over \epsilon_1 \epsilon_2 \beta }
\nonumber \\
&-{4N^3-6N^2+3N\over 24}{m^2(2\pi i-m)^2\over \epsilon_1 \epsilon_2 \beta}
+{ \mathcal{O}(\beta,\epsilon_{1,2}) \over \epsilon_1 \epsilon_2 \beta } .
\label{eq735_freE}
\end{empheq}
Since \(\epsilon_1 \epsilon_2 \beta<0\), the free energy \(\log Z\) is positive, which indicates the large number of degrees of freedom. Also, it is proportional to \(N^3\) at large \(N\) limit.

\par

\subsection{Asymptotic entropy}
In this section, we compute the free energy of the rank \(N\) E-string theory from the Cardy formula obtain in the previous subsection.
\par 
To simplify the computation, let us consider the case when \(Q_L=0\), so that \(Q_1=Q_2=Q_R\). We further take all \(E_8\) flavor charges to be the same \(F_l=F\), and define a chemical potential \(m_{E_8}=m_l\). Lastly, we shall turn off the tensor VEV \(\mathbf{v}\) to focus on the conformal phase. Then, the fugacity expansion of the \(\mathbb{R}^4\times T^2\) index can be written as follows,
\begin{align}
Z=\sum \Omega \cdot e^{-\mathfrak{n}\cdot \mathbf{v} } e^{-\epsilon_1(J_1+Q_R)-\epsilon_2(J_2+Q_R)}e^{-\beta P}e^{-8m_{E_8} F}
\end{align}
The summation is taken over all possible charges and the string number \(\mathfrak{n}\). Then, by following the same argument given in section \ref{section24}, the asymptotic entropy in the Cardy limit can be obtained from the following expression,
\begin{align}
S\simeq \log Z+\mathfrak{n}\cdot \mathbf{v}+\epsilon_1(J_1+Q_R)+\epsilon_2(J_2+Q_R)+\beta P+8m_{E_8}F
\label{eq863_ext}
\end{align}
which should be extremized with respect to the chemical potentials \(\epsilon_{1,2}\), \(\beta\), \(m_{E_8}\), and the tensor VEV \(v\).
\par 
The free energy is given in (\ref{eq791_freeenergy}), and we will assume that (\ref{eq791_freeenergy}) holmorphically extends to complex values of the chemical potentials inside a chamber \(-{\pi\over 2}<\text{Im}[m_{E_8}]<{\pi\over 2}\). In order to extremize (\ref{eq863_ext}), we should solve the following saddle point equations,
\begin{align}
{\partial S\over \partial \epsilon_1}=0,\quad 
{\partial S\over \partial \epsilon_2}=0,\quad
{\partial S\over \partial \beta}=0,\quad 
{\partial S\over \partial m_{E_8}}=0,\quad 
{\partial S\over \partial v_a}=0.
\end{align}

After solving the saddle point equations, the charges and the entropy are given as follows,
\begin{align}
&J_1+Q_R\simeq -{c_1 \pi^4+4\pi^2 c_2 (2m_{E_8}^2+\pi^2)+16 c_3(2m_{E_8}^2+\pi^2)^2 \over \epsilon_1^2 \epsilon_2 \beta}
\nonumber \\
&J_2+Q_R\simeq -{c_1 \pi^4+4\pi^2 c_2 (2m_{E_8}^2+\pi^2)+16 c_3(2m_{E_8}^2+\pi^2)^2 \over \epsilon_1 \epsilon_2^2  \beta}
\nonumber \\
&P\simeq -{c_1 \pi^4+4\pi^2 c_2 (2m_{E_8}^2+\pi^2)+16 c_3(2m_{E_8}^2+\pi^2)^2 \over \epsilon_1 \epsilon_2 \beta^2}
\nonumber \\
&F\simeq {2\pi^2 c_2 m_{E_8}+16 c_3 m_{E_8}(2m_{E_8}^2+\pi^2) \over \epsilon_1 \epsilon_2 \beta}
\nonumber \\
&n_a\simeq -(N+1-a)\cdot {2\pi^2+ {1\over 2}\sum_l m_l^2 \over \epsilon_1 \epsilon_2}- {N^2-N-2+3a-a^2\over 2}\cdot {m(m-2\pi i)\over \epsilon_1 \epsilon_2}
\nonumber \\
&S\simeq -{4\pi^4 c_1+16 \pi^2 c_2(m_{E_8}^2+\pi^2)+64\pi^2 c_3(2m_{E_8}^2+\pi^2) \over \epsilon_1 \epsilon_2 \beta}
\end{align}
where \(\simeq\) means the above expression hold at the leading order in the Cardy limit. Also, constants \(c_{1,2,3}\) are
\begin{align}
c_1={4N^3-6N^3+3N\over 24},\quad 
c_2={N^3-N\over 4},\quad 
c_3={N\over 8}.
\end{align}
Now, it is straightforward to express the entropy \(S\) in terms of the charges. However, the dependence of the flavor charge \(F\) is extremely complicated. Hence, here, we express the entropy as a series expansion of the flavor charge as follows,
\begin{align}
S&=2\pi \Bigr(-{2\over 3}(4N^3+18N^2+27N) (J_1+Q_R)(J_2+Q_R)P \Bigr)^{1\over 4}
\nonumber\\
&\times \Bigr( 1-\sqrt{4N^2+18N+27\over  6N^3+36N^2+54N} {F^2\over \sqrt{-(J_1+Q_R)(J_2+Q_R)P}}+\mathcal{O}(F^4) \Bigr).
\label{eq901_ent_ser}
\end{align}
The series expansion is valid when \(F^2\ll \sqrt{-(J_1+Q_R)(J_2+Q_R)P}\). One may wonder that such condition is not obviously guaranteed in the Cardy limit since they are at the same order, i.e. \(F^2\sim \sqrt{-(J_1+Q_R)(J_2+Q_R)P}\sim \mathcal{O}(\epsilon_1^{-2}\epsilon_2^{-2}\beta^{-2}) \). However, if we further consider the large \(N\) limit, one can check that \(J_{1,2}+Q_R\) and \(P\) are \(\mathcal{O}(N^3)\) while \(F\sim \mathcal{O}(N^2)\) if the flavor chemical potential \(m_{E_8}\) is \(\mathcal{O}(1)\). Then, the correction term in the second line is \(\mathcal{O}(N^{-1})\), and the series expansion of the form (\ref{eq901_ent_ser}) is valid in the large \(N\) limit.

\section{Background field analysis}
\label{section4}
In this section, we study the Cardy free energy by using the background field analysis on \(\mathbb{R}^4\times T^2\). We shall mainly follow the logic developed in \cite{Nahmgoong:2019hko}, which we will briefly explain here. The basic idea is to view the chemical potentials in the index as the background fields of the corresponding global symmetries. On \(\mathbb{R}^4\times T^2\), the chemical potentials \(\epsilon_{1,2}\) and \(\beta\) are reflected on the background metric as follows, 
\begin{align}
g_{\mu\nu} dx^\mu dx^\nu
&=\sum_{a=1}^2\Bigr(dz_a+{\epsilon_a\over  2\pi r \text{Im}[\tau] } z_a dy \Bigr)
\Bigr(d\bar{z}_a -{\epsilon_a\over 2\pi r \text{Im}[\tau] } \bar{z}_a  dy \Bigr)
+(dx-{\text{Re}[\tau] \over \text{Im}[\tau] }dy )^2+dy^2
\label{eq481_metric}
\end{align}
where \(z_a=x_a+iy_a\) are complex coordinates of \(\mathbb{R}^4\). Also, we shall set \(\epsilon_{1,2}\) and \(\tau\) to be complex chemical potentials in this section. The torus coordinate \(x\) and \(y\) have the following periodicity,
\begin{align}
(x,y)\sim (x+2\pi r ,y) \sim (x+2\pi r \text{Re}[\tau],y+2\pi r\text{Im}[\tau])
\label{eq959_period}
\end{align}
where \(x\) is the spatial circle, and \(y\) is the Euclidean time coordinate.  If there is another \(U(1)\) global symmetry chemical potential \(\Delta\), it is encoded in the background \(U(1)\) gauge field as follows,
\begin{align}
\mathbf{A}={\Delta\over 2\pi r \text{Im}[\tau] }dy
\label{eq965_bkcslsd}
\end{align}
where we used a bold-faced letter for 1-form fields.
\par 
Now, let us consider the \(\mathbb{R}^4\times T^2\) index. It can be obtained from the path integral over the 6d dynamic fields, with the supersymmetric boundary condition on the temporal circle. Then, we can consider a dimensional reduction of the 6d theory on the small temporal circle. Then, the heavy KK modes of the dynamic fields are integrated out, and we finally obtain the 5d effective field theory of the background fields of \(g_{\mu\nu}\) and \(\mathbf{A}\). It can be justified in the Cardy limit where \(|\tau| \ll  1\). In this limit, the temporal circle becomes much smaller than the spatial circle of the torus, and we obtain a 5d theory on a \(\mathbb{R}^4\times S^1\). 
\par 
Therefore, the \(\mathbb{R}^4\times T^2\) index \(Z\) can be obtained from the 5d effective action \(W_\text{eff}\) on \(\mathbb{R}^4\times S^1\) as \(\log Z \sim -W_\text{eff}\). Rigorously speaking, we have to consider the effect of dynamic zero-modes, which remain massless upon the KK compactification. However, the leading Cardy free energy is inversely proportional to the temporal circle radius, i.e., \(\log Z=\mathcal{O}(\tau^{-1})\), and we expect that the zero-mode contribution is absent at the leading order \cite{Choi:2018hmj}.
\par 
Now, let us consider the structure of the 5d effective action \(W_\text{eff}\). Upon the dimensional reduction, the 6d background fields (\ref{eq481_metric}) and (\ref{eq965_bkcslsd}) are decomposed into 5d fields as follows,
\begin{align}
g_{\mu\nu} dx^\mu dx^\nu =e^{-2\Phi}(dy+\mathbf{a})^2+h_{ij} dx^i dx^j,\quad 
\mathbf{A}=A_0 (dy+\mathbf{a})+\hat{\mathbf{A}}
\end{align}
where \(\Phi\) is a dilton, \(\mathbf{a}\) is a gravi-photon, \(h_{ij}\) is a 5d metric, \(A_0\) is a temporal scalar, and \(\hat{\mathbf{A}}\) is a 5d background gauge field. In \(W_\text{eff}\), there are infinitely many terms arranged in the derivative expansion, and the exact expression cannot be without evaluating the path integral honestly. However, among the infinitely many terms in the derivative expansion, the 5d Chern-Simons terms can be determined from the 't Hooft anomalies of the 6d theory \cite{DiPietro:2014bca,Kim:2017zyo}.
\par 
Recall that the 't Hooft anomalies determine the non-invariance of the effective action with respect to the background field transformation. In the 5d CS terms, this non-invariance is encoded in the non-invariant CS terms, which are the normal CS terms multiplied with the temporal scalars, such as \(A_0 \hat{\mathbf{A}}d\hat{\mathbf{A}}d\hat{\mathbf{A}}\). On the other hand,  the invariant CS terms such as \(\hat{\mathbf{A}}d\hat{\mathbf{A}}d\hat{\mathbf{A}}\) do not contribute to the non-invariance of the effective action, but they are correlated with the non-invariant CS terms in a complicated manner \cite{DiPietro:2014bca, Chang:2019uag}. For example, in \cite{Jensen:2012kj, Jensen:2013kka, Jensen:2013rga}, the structure of the invariant CS terms is completely determined from the non-invariant CS terms by demanding the `consistency condition with the Euclidean vacuum.' 
\par 
Knowing the CS terms in the effective action is indeed important since they determine the leading order behavior of the effective action in the Cardy limit \cite{Choi:2018hmj, Nahmgoong:2019hko}. For the 6d (2,0) SCFTs and the rank \(N\) E-string theory, the CS terms in the 5d effective action were computed in \cite{Nahmgoong:2019hko} using the consistency condition with the Euclidean vacuum. In the following subsections, we will use those effective action to test our Cardy formulas (\ref{eq343_cardy}) and (\ref{eq735_freE}).

\subsection{6d (2,0) SCFT}
Let us first consider the 6d (2,0) SCFTs of ADE type. In order to study the index from the background field method, we consider the \(\mathbb{R}^4\times T^2\) index in the following modified basis,
\begin{align}
\mathcal{Z} (\tau,\epsilon_1,\epsilon_2,\Delta_1,\Delta_2)=\text{Tr}\Bigr[ e^{-\beta P} e^{-\epsilon_1 J_1-\epsilon_2 J_2} e^{-\Delta_1 Q_1-\Delta_2 Q_2} \Bigr]
\label{eq295_modind}
\end{align}
with the chemical potential constraint \(\Delta_1+\Delta_2-\epsilon_1 -\epsilon_2=2\pi i\) imposed. The index in this basis is also known as the `modified index' \cite{Kim:2019yrz}. Here, \(Q_{1,2}\)'s are two Cartans of R-symmetry, which rotate two orthogonal planes in \(SO(5)_R\). They are normalized to be an integer for bosonic states and half-integer for fermionic states, i.e., \((-1)^F=e^{-2\pi i Q_2}\). As a result, the modified index (\ref{eq295_modind}) can be turned into the original index (\ref{eq48_index}) by the following chemical potential mapping,
\begin{align}
\mathcal{Z} (\tau,\epsilon_1,\epsilon_2,\Delta_1=\epsilon_+ +m,\Delta_2=2\pi i+\epsilon_+ -m)
=Z(\tau,\epsilon_1,\epsilon_2,m).
\label{eq350_modind}
\end{align}
\par 
Now, let us consider the following Cardy limit,
\begin{align}
|\epsilon_{1,2}|, \ |\tau| \ll 1.
\end{align}
As we have discussed, the leading order behavior of free energy in the Cardy limit is given by the CS terms in the 5d effective action. Also, the CS terms can be determined from the 't Hooft anomaly. The anomaly polynomial of the 6d (2,0) SCFT given as follows,
\begin{align}
I_8={h_G^\vee d_G\over 24}p_2(N)+{r_G\over 48}\Bigr[ p_2(N)-p_2(T)+{1\over 4}\Bigr(p_1(N)-p_1(T)\Bigr)^2 \Bigr]
\end{align}
where \(h_G^\vee\) is a dual Coxeter number, \(d_G\) is a dimension, and \(r_G\) is a rank of a Lie algebra \(G\in\{A,D,E\}\). Also, \(p_{1,2}(N/T)\) are the Pontryagin classes of \(SO(5)_R\) normal bundle and \(SO(4)\) tangent bundle. The corresponding 5d CS terms in the effective action were determined in \cite{Nahmgoong:2019hko}. It takes the following form\footnote{We flipped the overall sign of \(W_\text{CS}\) in \cite{Nahmgoong:2019hko} due to the opposite 6d chirality convention. When defining the \(\mathbb{R}^4\times T^2\) index in this paper, we used the anti-chiral supercharges on \(\mathbb{R}^4\) which are (0,2) spinors in the convention of \cite{Nahmgoong:2019hko}. },
\begin{align}
W_\text{CS}&={i\over  (2\pi  r \text{Im}[\tau])^3}\int_{\mathbb{R}^4\times S^1}\mathbf{a}d\mathbf{a}d\mathbf{a} \Bigr( {h_G^\vee d_G+r_G\over 3072\pi^3}(\Delta_R^2-\Delta_L^2)^2
+{r_G\over 1536\pi^3} (\Delta_R^2+4\pi^2)(\Delta_L^2+4\pi^2)\Bigr)
\nonumber \\
&+W_\text{mixed-CS}+W_\text{grav-CS}.
\label{eq382_20effective}
\end{align}
where \(\Delta_R\equiv \Delta_1+\Delta_2\) and \(\Delta_L\equiv \Delta_1-\Delta_2\). Here, \(W_\text{mixed-CS}\) contains the mixed CS terms between the gauge the the gravitational fields, and \(W_\text{grav-CS}\) contains the pure gravitational CS terms. However, they are subleading in the Cardy limit \cite{Nahmgoong:2019hko}, and we shall focus on the gauge CS terms only. From the 6d metric (\ref{eq481_metric}), one can obtain the 5d graviphoton field as follows,
\begin{align}
\mathbf{a}={2\pi r \text{Im}[\tau] \over \epsilon_1^2 |z_1|^2+\epsilon_2^2 |z_2|^2+4\pi^2 r^2 |\tau|^2 }\Bigr( \epsilon_1 (-y_1 dx_1+x_1 dy_1) +\epsilon_2 (-y_2 dx_2+x_2 dy_2)-2\pi r \text{Re}[\tau] dx \Bigr)
\end{align}
Then the graviphoton CS action is given as follows,
\begin{align}
{1\over (2\pi  r \text{Im}[\tau])^3} \int_{\mathbb{R}^4\times S^1} \mathbf{a} d\mathbf{a}d\mathbf{a}
={4\pi^2 \over \epsilon_1 \epsilon_2} {\text{Re}[\tau]\over |\tau|^2}
={(2\pi)^3 \over \epsilon_1 \epsilon_2}\text{Im}\Bigr[ {1\over \beta } \Bigr].
\label{eq324_adada}
\end{align}
Inserting (\ref{eq324_adada}) to (\ref{eq382_20effective}), we obtain the following CS contribution in the effective action,
\begin{align}
-W_\text{CS}=\Bigr(-{h_G^\vee d_G+r_G\over 384 } {(\Delta_R^2-\Delta_L^2)^2 \over \epsilon_1 \epsilon_2  } 
-{r_G\over 192}  {(\Delta_R^2+4\pi^2)(\Delta_L^2+4\pi^2) \over \epsilon_1 \epsilon_2 }\Bigr)\cdot i\text{Im}[{1\over \beta}]
+{ \mathcal{O}(\beta,\epsilon_{1,2}) \over \epsilon_1 \epsilon_2 \beta } 
\end{align}
where \(\Delta_{R,L}\) are set to be purely imaginary and the subleading \(\mathcal{O}(\tau^{-2})\) term includes contributions from \(W_\text{mixed-CS}\) and \(W_\text{grav-CS}\). From the index condition (\ref{eq350_modind}), we insert \(\Delta_R=2\pi i +2\epsilon_+\) and \(\Delta_L=-2\pi i +2m\). Then the effective action becomes, 
\begin{align}
-W_\text{CS}= -{h_G^\vee d_G+r_G\over 24 } {m^2(2\pi i -m)^2\over \epsilon_1 \epsilon_2  } \cdot i\text{Im}[{1\over \beta}]+{ \mathcal{O}(\beta,\epsilon_{1,2}) \over \epsilon_1 \epsilon_2 \beta } .
\label{eq530_CSIM}
\end{align}
For \(A_{N-1}\) theory, \(h_G^\vee d_G+r_G=N^3-1\). Therefore, considering the one free tensor multiplet contribution, the CS terms in the background effective action exactly reproduce the imaginary part of the Cardy free energy (\ref{eq343_cardy}). We expect that the supersymmetric completion of the CS terms will recover the holomorphic dependence of \(\tau\), which will generate the real part of the free energy also.

\subsection{E-string theory}
As the other example, let us consider the rank 1 E-string theory. Let us define the \(\mathbb{R}^4\times T^2\) index in the following modified basis,
\begin{align}
\mathcal{Z}=\text{Tr}\Bigr[  e^{-\beta  P} e^{-\epsilon_1 J_1-\epsilon_2 J_2} e^{-\Delta_R Q_R }e^{-\Delta_L Q_L}\prod_{l=1}^8 e^{-m_l F_l} \Bigr]
\label{eq352_modEstring}
\end{align}
where the chemical potentials are constrained by \(\Delta_R-\epsilon_1-\epsilon_2=2\pi i\). Here, \(Q_R\) is the Cartan of \(SU(2)_R\) R-symmetry, and \(Q_L\) is the Cartan of \(SU(2)_L\) flavor symmetry. Since \(SU(2)_R\times SU(2)_L\sim SO(4)\), \(Q_R\pm Q_L\) is an integer for bosonic states and half-integer for fermionic states, i.e., \(e^{-2\pi i (Q_R \pm Q_L)}=(-1)^F\). Therefore, one can turn the modified index (\ref{eq352_modEstring}) into the original index (\ref{eq216_Estring}) by the following chemical potential mapping,
\begin{align}
\mathcal{Z}(\tau,\epsilon_1,\epsilon_2,\Delta_R=2\pi i +2\epsilon_+,\Delta_L=-2\pi i+2m,m_l)
=Z (\tau,\epsilon_1,\epsilon_2,m_l).
\end{align}
\par 
The anomaly polynomial of the rank \(N\) E-string theory is given in (\ref{eq660_anom_Estring}). The corresponding CS terms in the background effective are given by\footnote{We flipped the overall sign of \(W_\text{CS}\) in \cite{Nahmgoong:2019hko} due to the opposite 6d chirality convention. }
\begin{align}
W_\text{CS}&=-{i\over 256\pi^3 \beta^3 }\int_{\mathbb{R}^4\times S^1}\mathbf{a}d\mathbf{a}d\mathbf{a} \Bigr[ -{N^3\over 3} (\Delta_R^2-\Delta_L^2)^2
+{N^2\over 2} (\Delta_R^2-\Delta_L^2)(4\sum_l m_l^2+8\pi^2-\Delta_R^2-\Delta_L^2)
\nonumber \\
&-{N\over 12}\Bigr(3(\Delta_R^4+\Delta_L^4)+8\Delta_R^2 \Delta_L^2-40\pi^2 (\Delta_R^2+\Delta_L^2) 
+24(8\pi^2-\Delta_R^2-\Delta_L^2)\sum_l m_l^2
\nonumber \\
&+48 (\sum_l m_l^2)^2+224\pi^4 \Bigr) \Bigr] +W_\text{mixed-CS}+W_\text{grav-CS}
\end{align}
Now, we use (\ref{eq324_adada}) to evaluate the integral. After inserting \(\Delta_R=2\pi i +2\epsilon_+\), \(\Delta_L=-2\pi i+2m\), we obtain the following result,
\begin{align}
-W_\text{CS}=&\Bigr[-{N\over 8}{(4\pi^2+ \sum_l m_l^2)^2\over \epsilon_1 \epsilon_2 }-{N(N-1)\over 4}{(4\pi^2+\sum_l m_l^2)m(m-2\pi i)\over \epsilon_1 \epsilon_2  }
\nonumber \\
&-{4N^3-6N^2+3N\over 24}{m^2(2\pi i-m)^2\over \epsilon_1 \epsilon_2 }\Bigr]\cdot i \text{Im}[{1\over \beta}]+{ \mathcal{O}(\beta,\epsilon_{1,2}) \over \epsilon_1 \epsilon_2 \beta }.
\end{align}
As expected, the CS terms reproduce the imaginary part of the Cardy free energy (\ref{eq735_freE}).

 \section{Concluding remarks}
 \label{section5}
In this paper, we computed the free energy of (2,0) \(A\)-type theory and the rank \(N\) E-string theory on \(\mathbb{R}^4\times T^2\) in the Cardy limit. For the SCFT on \(N\) M5-branes, the Cardy free energy is proportional to \(N^3\) in the conformal phase, where the tensor VEVs are small. If we increase the tensor VEVs, we can observe the phase transition on the tensor branch, and the free energy is reduced to the Abelian contribution proportional to \(N^1\). The number of the self-dual string has non-zero expectation value at the conformal phase, and it plays a vital role in understanding \(N^3\) degrees in 6d SCFTs. The distribution of the self-dual string is identical to the Weyl vector of \(A_{N-1}\) Lie algebra, and our result is consistent with the D0-D4 bound state counting \cite{Kim:2011mv}. Notably, the total number of the self-dual string is proportional to \({N^3-N\over 6}\), which is proportional to the non-Abelian part of the Cardy free energy. The detailed picture of the $N^3$ degrees of freedom is more complicated as argued at in section \ref{section22}. These strings would carry momentum and other quantum numbers while also forming threshold bound states. We hope to come back to this problem and clarify the microscopic physics of those degrees in the future.

\par 

Our analysis is extended to (1,0) E-string theory of rank \(N\). We obtain the free energy, which is also proportional to \(N^3\) in the large \(N\) limit. Also, we show that the number of the self-dual strings have non-zero expectation value, whose distribution is more complicated than (2,0) A-type theory. The self-dual strings are more condensed near the M9-brane, and it is not described by the Weyl vector of any Lie algebra. However, the total number of the self-dual strings is again proportional to \(N^3\) in the large \(N\) limit. 

\par 

The condensation of the self-dual strings in the Cardy limit is important to study the deconfining degrees in 6d SCFTs. In general, we speculate that the string condensation is closed related to the inverse of the intersection matrix \(\Omega\), which becomes a `quadratic form' matrix of Lie algebra in the case of (2,0) theories. More precisely, the string in the \(a\)'th tensor branch has the VEV \(\langle n_a \rangle\) proportional to \(\sum_b \Omega^{-1}_{a,b}\). It would be interesting to perform a similar computation to (2,0) D/E-type theories and many other (1,0) theories. 

 \par 
Our results in this paper are focused on the \(\mathbb{R}^4\times T^2\) index, and especially its elliptic genus expansion. However, there seem to be many future directions worth pursuing. We finish this paper by listing some of them. 
 \begin{itemize}
 \item  DLCQ index in the Cardy limit
 \end{itemize}
 \par 
On the tensor branch moduli space, the origin \(\mathbf{v}=0\) is a special point where the conformal symmetry is restored. In the M-theory brane set-up, it corresponds to the setting where all M5-branes coincide. However, it was found that the tensor branch index on \(\mathbb{R}^4\times T^2\) becomes divergent if we set \(\mathbf{v}=0\) \cite{Kim:2011mv}. Instead, as a conformal phase observable, one can study the discrete light-cone quantization (DLCQ) index of the 6d SCFT. The DLCQ description of the 6d (2,0) theory was studied in \cite{Aharony:1997th, Aharony:1997an} by compactifying it on a null circle of the light cone. The subgroup of 6d superconformal symmetry is preserved, which commutes with the light cone momentum.  The DLCQ index of 6d SCFTs was studied in \cite{Kim:2011mv} for (2,0) \(A\)-type theories, and it can be obtained by projecting gauge singlets in the 5d instanton partition function as follows,
 \begin{align}
 Z_\text{DLCQ}=\sum_{k=0}^\infty q^k
 \oint  [\prod_{a=1}^N {d(e^{-v_a})\over e^{-v_a}}] \cdot \Bigr( \prod_{\alpha } 2\sinh{\alpha\cdot \mathbf{v} \over 2}\Bigr) \cdot Z_k.
 \label{eq935_CLDQ}
 \end{align}
Here, \(Z_k\) is the \(k\)-instanton partition function of 5d \(U(N)\) MSYM and \(\alpha \in \text{root}(A_{N-1})\). It was further check that the index (\ref{eq935_CLDQ}) is equivalent to the DLCQ supergraviton index of dual \(AdS_7\) in the large \(N\) limit \cite{Kim:2011mv}.
\par 
Therefore, studying the behavior of the DLCQ index in the Cardy limit is essential to better understand the 6d SCFTs in the conformal phase. Also, the DLCQ Cardy formula might imply the possible existence of the black holes in dual \(AdS_7\). In this paper, our approach to obtain the free energy in the Cardy limit is smooth when we set \(\mathbf{v}\to 0\). The divergence at the conformal point \(\mathbf{v}=0\) is invisible in our Cardy free energy since its effect is subleading in the Cardy limit. Therefore, it seems interesting if we can infer the Cardy free energy of the DLCQ index from our result at \(\mathbf{v}\to 0\). 

\begin{itemize}
\item Superconformal index and \(AdS_7\) black holes
\end{itemize}

Recently, there has been a substantial development of counting the entropy of \(AdS_5\) black holes from the superconformal index 4d \(\mathcal{N}=4\) SYM on \(S^3\times S^1\) \cite{Cabo-Bizet:2018ehj,Choi:2018hmj, Benini:2018ywd} in the large \(N\) limit. It was further generalized into other dimensions. For \(AdS_7/CFT_6\), the entropy of BPS black holes in \(AdS_7\) was counted from the dual 6d SCFTs using the background effective action \cite{Nahmgoong:2019hko} and the Casimier energy \cite{Kantor:2019lfo}. However, it is still less understood how the state-counting part of the 6d superconformal index accounts for the black hole entropy. 
\par 
In this paper, we focused on the tensor branch index on \(\mathbb{R}^4 \times T^2\). It is different from the superconformal index in various points, but it is the ingredient to compute the superconformal index on \(S^5\times S^1\). Therefore, let us make a few comments on the implication of our results to the superconformal index and \(AdS_7\) black holes.
\par 
The 6d (2,0) SCFT on \(S^5\times S^1\) has \(SO(6)\) angular momenta \(J_{1,2,3}\) and \(SO(5)\) R-charges \(Q_{1,2}\). The superconformal index is defined as \cite{Bhattacharya:2008zy}
\begin{align}
Z_{S^5\times S^1} =\text{Tr}[(-1)^F e^{-\omega_1 J_1-\omega_2 J_2-\omega_3 J_3}e^{-\Delta_1 Q_1-\Delta_2 Q_2}]
\end{align}
where the chemical potentials are constrained by \(\Delta_1+\Delta_2-\omega_1-\omega_2-\omega_3=0\). The superconformal index uses the three copies of the tensor branch indices as the integrand whose chemical potentials are given by \((\beta,\epsilon_1,\epsilon_2,m)=(\omega_1,\omega_{21},\omega_{31},{\Delta_L-\omega_1\over 2})\), \((\omega_2,\omega_{12},\omega_{32},{\Delta_L -\omega_2\over 2})\), and \((\omega_3,\omega_{13},\omega_{23},{\Delta_L-\omega_3\over 2})\) respectively, where \(\omega_{ij}=\omega_i-\omega_j\) and \(\Delta_L=\Delta_1-\Delta_2\). The relevant formula can be found in \cite{Kim:2013nva}. If we plug the chemical potential mapping to our \(\mathbb{R}^4\times T^2\) Cardy formula at the conformal phase, we obtain the following result,
\begin{align}
\log Z_{S^5\times S^1}&\sim \Bigr[-{N^3\over 24}{m^2(2\pi i-m)^2\over \beta \epsilon_1 \epsilon_2} \Bigr]_{(\beta,\epsilon_1,\epsilon_2,m)=(\omega_1,\omega_{21},\omega_{31},{\Delta_L-\omega_1\over 2})}+\text{permutations}
\nonumber \\
&=-{N^3\over 24}{ \hat{\Delta}_1^2 \hat{\Delta}_2^2 \over \omega_1 \omega_2 \omega_3}+\mathcal{O}(\omega^{-2})
\end{align}
where \(\hat{\Delta}_1=\Delta_1\) and \(\hat{\Delta}_2=\Delta_2+2\pi i\), so that \(\hat{\Delta}_1+\hat{\Delta}_2-\omega_1-\omega_2-\omega_3=2\pi i\). The result is the expected free energy of the 6d SCFT on \(S^5\times S^1\) in the Cardy limit \(|\omega_i|\ll 1\) \cite{Nahmgoong:2019hko}. Also, it reproduces the Bekenstein-Hawking entropy of large BPS black holes in \(AdS_7\times S^4\) \cite{Hosseini:2018dob}. We hope to rigorously derive the superconformal Cardy formula from the localization on \(S^5\times S^1\) in the near future.

 \begin{itemize}
 \item  Cardy formula from the instanton partition functions
 \end{itemize}
 \par 
For the 6d SCFTs on \(\mathbb{R}^4\times T^2\), there are two parallel approaches which are based on the elliptic genus expansion (\ref{eq65_index2}) and the instanton partition function expansion (\ref{eq151_pertinst}). It would be important if we can obtain the 6d Cardy formula on \(\mathbb{R}^4\times T^2\) from the instanton partition function of 5d maximal SYM on \(\mathbb{R}^4\times S^1\). In (\ref{eq626_quntities}), we showed that the KK momentum, which becomes the instanton number, scales as \(\mathcal{O}({1\over \epsilon_1 \epsilon_2 \beta^2})\) in the Cardy limit. Therefore, it is an interesting question to ask whether instantons also condensate similar to the self-dual strings, and the condensation is identical to our prediction in (\ref{eq626_quntities}).

\par

Investigating the Cardy formula from the instanton partition function might also become a useful approach for the 5d SCFTs on \(\mathbb{R}^4\times S^1\). For the 5d superconformal indices on \(S^4\times S^1\), the instanton partition function is suppressed, and it does not contribute to the \(\mathcal{O}(N^{5\over 2})\) free energy \cite{Choi:2019miv}. However, if the 5d free energy also scales as \(\mathcal{O}(N^{5\over 2})\) for the Coulomb branch on \(\mathbb{R}^4\times S^1\), it cannot be reproduced with the perturbative partition function only. Hence, we expect that studying 5d Cardy formulas on \(\mathbb{R}^4\times S^1\) from the instanton partition function can shed light on the deconfining mechanism of \(N^{5\over 2}\) degrees of freedom.

\par 

As a final remark, we speculate that the Gopakumar-Vafa (GV) invariant \cite{Gopakumar:1998jq, Iqbal:2007ii} can also become a useful observable to understand 5d/6d SCFTs in the Cardy limit. The GV invariant counts the number of BPS states, and it encodes the same information with the Nekrasov partition function. Also, the Cardy limit physics can be studied from the GV invariants in the large spin limit. There is an upper bound on the spins called the `Castelnuovo bound' \cite{Gu:2017ccq} such that the GV invariants vanish beyond it. However, the asymptotics of the GV invariants in the large spin limit is less studied. Therefore, studying the Cardy formula from the GV invariant can also become a novel approach to understand the Cardy limit physics from a different perspective.

\acknowledgments
We thank Seok Kim for helpful comments and inspiring discussions. KL is supported by   KIAS Individual Grant  PG006904 and National Research Foundation Individual Grant NRF-2017R1D1A1B06034369. JN is supported by KIAS Individual Grant PG076401.


\bibliographystyle{JHEP}
\bibliography{main}

\end{document}